# Efficient update of redundancy matrices for truss and frame structures

Tim Krake[1,2], Malte von Scheven[3], Jan Gade[3], Moataz Abdelaal[1],
Daniel Weiskopf[1], and Manfred Bischoff[3]

[1] University of Stuttgart, Visualization Research Center (VISUS)
[2] Hochschule der Medien
[3] University of Stuttgart, Institute for Structural Mechanics

Redundancy matrices provide insights into the load carrying behavior of statically indeterminate structures. This information can be employed for the design and analysis of structures with regard to certain objectives, for example reliability, robustness, or adaptability. In this context, the structure is often iteratively examined with the help of slight adjustments. However, this procedure generally requires a high computational effort for the recalculation of the redundancy matrix due to the necessity of costly matrix operations. This paper addresses this problem by providing generic algebraic formulations for efficiently updating the redundancy matrix (and related matrices). The formulations include various modifications like adding, removing, and exchanging elements and are applicable to truss and frame structures. With several examples, we demonstrate the interaction between the formulas and their mechanical interpretation. Finally, a performance test for a scaleable structure is presented.

## 1 Introduction

The redundancy matrix is important for characterizing load carrying structures. It describes the distribution of internal constraint and thus the ability of statically indeterminate structures to carry loads along different paths. The degree of statical indeterminacy is a system-inherent property of the spatially discrete, elastostatic models of a truss or frame structure. It is defined as the difference between the number of unknown force quantities and the number of linearly independent equilibrium equations. Usually, statical indeterminacy is only considered as one aggregated integer number characterizing the entire structure or substructures. In contrast, the redundancy matrix also provides the spatial distribution of the degree of statical indeterminacy as well as information about different load carrying mechanisms. This additional information yields important insights into the load carrying behavior for many aspects of design and analysis.

Driven by the analogy between Gaussian adjustment calculus in geodesy and structural mechanics, a concept of redundancy in statically indeterminate truss and frame structures was developed by Linkwitz (1961); Bahndorf (1991); Ströbel (1997) and revisited by von Scheven et al. (2021). Another more recent work on the redundancy matrix was published by Liu and Liu (2005), who used the redundancy matrix and its counterpart, the 'importance indices', to assess the importance of structural elements in truss and frame structures. The idea of the redundancy matrix was extended to kinematically indeterminate structures (Tibert 2005; Zhou et al. 2015; Chen et al. 2018) and to continuous representations of structures by a connection between concepts from applied mathematics and classical structural mechanics (Gade et al. 2021). Further references for the calculation and application of the redundancy matrix can be found in the works by von Scheven et al. (2021) and Gade et al. (2021).

The redundancy matrix can be used in numerous applications, including reliability and robust design of structures (Frangopol and Curley 1987; Pandey and Barai 1997; Kou et al. 2017;

Spyridis and Strauss 2020), quantification of imperfection sensitivity (Eriksson and Tibert 2006; Ströbel and Singer 2008), assessment of adaptability as well as actuator placement, and optimized control in adaptive structures (Wagner et al. 2018; Geiger et al. 2020; Maierhofer and Menges 2019). However, this paper focuses on the efficient update procedures for the redundancy matrix while the mechanical interpretation of the redundancy distribution can be found in the references given above.

During the design process of structures, many different variants are usually studied. In this case, the redundancy distribution in the structure can be used as an indicator of the load bearing behaviour or robustness of the different variants. These values can be either used as direct feedback to interactive design changes by an engineer or in an automatic optimization process (Bahndorf 1991). In both cases, a fast calculation of the redundancy matrix is essential. However, especially for larger structures, the recalculation of the redundancy matrix is time-consuming as it involves the inverse of a matrix whose dimensionality is given by the number of degrees of freedom in the system. It would be of great advantage to replace the recalculation by an update of the redundancy matrix to reduce computational effort and enable immediate feedback to the engineer or faster optimization loops.

Chen et al. (2010) derive such an update formula for the redundancy matrix for removing single elements from a truss structure. The approach is used for the assessment of the reliability of structures and the redistribution of inner forces due to damage of components. The same formula is also applied to a truss system by Kou et al. (2017). However, this approach is limited to removing elements from truss structures and cannot be extended to other modifications of the structure, like adding elements or exchanging elements.

The Woodbury formula (Guttman 1946; Woodbury 1950), also known as the generalized Sherman-Morrison-Woodbury formula, provides an explicit formula for the inverse of a matrix after a rank-$k$ update. In the context of structural mechanics, it was applied to compute a modified flexibility matrix, e.g. after removal of elements (Sack et al. 1967; Argyris et al. 1971; Bahndorf 1991). A good overview of further references for the application of the Sherman-Morrison-Woodbury formula for structural reanalysis is provided by Akgün et al. (2001).

In case of a modification of the structural topology, the cross sections, or the elastic constants, the inverse of the modified stiffness matrix is required to compute the modified redundancy matrix. Thus, the computational complexity for the (re)computation of the modified redundancy matrix scales cubically with the size of the problem. Therefore, for systems with a large number of degrees of freedom an efficient update of the inverse elastic stiffness matrix is crucial to efficiently update the redundancy matrix. This motivates employing the Woodbury formula to the inversion of the elastic stiffness matrix. By this, the computationally expensive recalculation of the redundancy matrix can be replaced by an efficient update for successive modifications. In detail, we propose

- the first generic algebraic formulations for efficient updates of the redundancy matrix for various modifications like adding, removing, and exchanging elements,
- application of these generic formulations to truss and frame structures, and
- a quantitative analysis of accuracy and performance of the update processes.

Due to the general matrix notation used for the derivation, the updates can also be applied to groups of elements as well as to individual load-carrying types of elements. As the presented update formulations only involve matrix-vector operations, the computational complexity scales only quadratically with the problem size. Therefore, they prove to be very efficient and reduce the computational effort, especially for large-scale structures. In sum, our novel formulations provide the first framework to update the redundancy matrix in an efficient and generic way.

In the following Section 2, all relevant aspects of matrix structural analysis, the definition of the redundancy matrix, and the Woodbury formula to update the inverse after a rank-$k$ update are provided. In Section 3, the efficient updates of the redundancy matrix are derived for adding, removing, and exchanging elements and presented in ready-to-use algorithms. The first example in Section 4 shows the application and algebraic correctness of these algorithms for a small introductory example. A possible area of usage is demonstrated in the second example, while performance tests are shown in the last subsection, including a scalabiblity analysis. The paper concludes with a summary and an outlook.

## 2 Background

In this section, relevant aspects of matrix structural analysis as well as the concept of redundancy matrices are summarized. Moreover, the Woodbury formula, is the basis for the efficient update.

### 2.1 Matrix structural analysis

For discrete models of spatial truss and frame structures, static analysis can be represented in matrix notation (Argyris and Scharpf 1969; Przemieniecki 1968). It is described in the following.

Given is a discrete model consisting of $n$ degrees of freedom, $n_n$ nodes, and $n_e$ elements, each of which carries loads via $n_m$ load-carrying modes. The number of load-carrying modes is $n_m = 6$ for a spatial beam element, $n_m = 3$ for a plane beam element, and $n_m = 1$ for a plane or spatial truss element. In general, the model can consist of a combination of truss and beam elements, i.e., $n_m$ may vary between the elements. Therefore, the index $n_q$ is introduced that defines the number of all load-carrying modes of all elements ($n_q = n_m n_e$ in case of a pure truss or beam model).

The diagram shown in Figure 1 gives an overview of the relevant equations and quantities in matrix structural analysis for linear elastostatics. In this regard, the vector of generalized displacements $\mathbf{d} \in \mathbb{R}^n$ assigned to the nodes satisfying static equilibrium, material law, and compatibility can be computed by

$$\mathbf{A}^\top \mathbf{C} \mathbf{A} \mathbf{d} = \mathbf{f} + \mathbf{A}^\top \mathbf{C} \mathbf{e}_0. \tag{1}$$

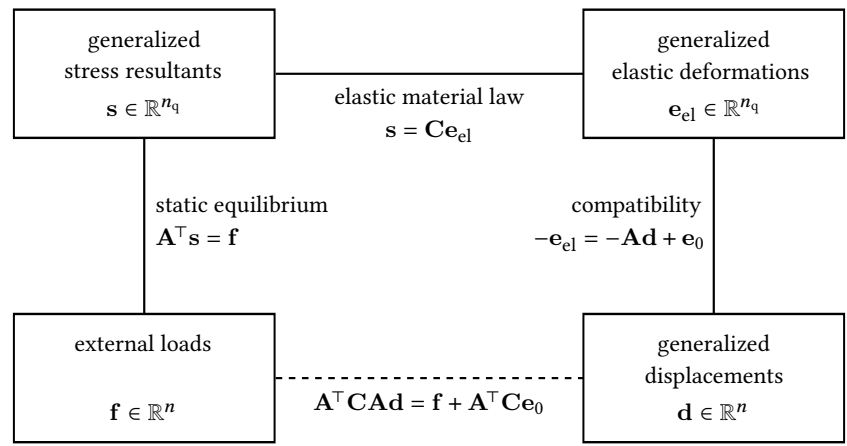

**Figure 1** Overview of relevant equations and quantities in matrix structural analysis for linear elastostatics (inspired by Tonti's diagram for elastostatic problems (Tonti 1976, p. 50) and by Strang (1986, p. 91)).

In Equation (1), $\mathbf{A}^\top \in \mathbb{R}^{n \times n_q}$ is the equilibrium matrix, $\mathbf{A} \in \mathbb{R}^{n_q \times n}$ is the compatibility matrix, and $\mathbf{C} \in \mathbb{R}^{n_q \times n_q}$ is the material matrix, which is a diagonal matrix with positive entries. The vector $\mathbf{f} \in \mathbb{R}^n$ represents the external loads and the vector $\mathbf{e}_0 \in \mathbb{R}^{n_q}$ represents the generalized pre-deformations. The matrix

$$\mathbf{K} = \mathbf{A}^\top \mathbf{C} \mathbf{A} \in \mathbb{R}^{n \times n} \tag{2}$$

is called the elastic stiffness matrix. It is symmetric by definition due to the diagonality of $\mathbf{C}$. Equation (1) can alternatively be obtained by minimizing the potential energy.

It is assumed throughout the paper that the structures are statically indeterminate with a degree of statical indeterminacy $n_s = n_q - \mathrm{rank}(\mathbf{A}^\top)$. Furthermore, it is assumed that the structures are kinematically determinate, i.e., $\mathrm{rank}(\mathbf{A}) = n$ (Pellegrino and Calladine 1986; Pellegrino 1993), which is equivalent to $\mathbf{K}$ being regular. The latter assumption can be satisfied by properly choosing structural topology and boundary conditions. It ensures that the structures are able to equilibrate loads without pre-stress (and thus geometric stiffness effects) such that linear structural theory is applicable.

## 2.2 Redundancy distribution

Based on the matrix structural analysis of the previous subsection, the concept of the redundancy (Bahndorf 1991; Ströbel 1997; von Scheven et al. 2021) is described in the following.

The redundancy distribution is independent of the external loads, thus $\mathbf{f} = \mathbf{0}$ is assumed. Solving Equation (1) for $\mathbf{d}$ and inserting into the compatibility equation (see Figure 1) yields

$$-\mathbf{e}_{\text{el}} = (\mathbf{I} - \mathbf{A}\mathbf{K}^{-1}\mathbf{A}^\top \mathbf{C})\mathbf{e}_0 = \mathbf{R}\mathbf{e}_0, \tag{3}$$

with the redundancy matrix

$$\mathbf{R} = \mathbf{I} - \mathbf{A}\mathbf{K}^{-1}\mathbf{A}^\top \mathbf{C} \in \mathbb{R}^{n_q \times n_q}. \tag{4}$$

Its main-diagonal entries $\mathbf{R}_{ii}$ provide the spatial distribution of the degree of statical indeterminacy of the structure (Bahndorf 1991; Ströbel 1997) such that $\sum_i^{n_q} \mathbf{R}_{ii} = n_s$. Due to the inverse of the stiffness matrix and the matrix-matrix multiplications, the computational complexity for the calculation of the redundancy matrix is given by $\mathcal{O}(n \cdot n_q^2)$. Since $n$ is typically proportional to $n_q$, the complexity scales cubically with the problem size.

A comparison of Equation (3) and the compatibility equation in Figure 1 reveals that $\mathbf{R}$ extracts the incompatible, stress-inducing parts from the generalized pre-deformations $\mathbf{e}_0$ taking the structural stiffness into account. In a statically determinate structure, $\mathbf{R} = \mathbf{0}$ holds; thus, it is $\mathbf{e}_{\text{el}} = \mathbf{0}$ for any $\mathbf{e}_0 \neq \mathbf{0}$. The redundancy matrix has several algebraic and spectral-theoretic properties. For more details, we refer to von Scheven et al. (2021) and Gade et al. (2021).

## 2.3 Woodbury formula

The Woodbury formula (Guttman 1946; Woodbury 1950) offers an explicit way to update a matrix after a rank-$k$ update (the related Sherman-Morrison-Woodbury formula characterizes the special case of rank-1 updates). To be more precise, let $\mathbf{M} \in \mathbb{R}^{n \times n}$ be an invertible matrix and $\mathbf{U}, \mathbf{V} \in \mathbb{R}^{n \times k}$ two other matrices such that $\mathbf{I} - \mathbf{V}^\top \mathbf{M}^{-1} \mathbf{U} \in \mathbb{R}^{k \times k}$ is invertible. Then, the following assertion holds:

$$(\mathbf{M} - \mathbf{U}\mathbf{V}^\top)^{-1} = \mathbf{M}^{-1} + \mathbf{M}^{-1}\mathbf{U}(\mathbf{I} - \mathbf{V}^\top \mathbf{M}^{-1}\mathbf{U})^{-1}\mathbf{V}^\top \mathbf{M}^{-1}. \tag{5}$$

We will use this formula to derive our algebraic formulations for efficient updates.

## 3 Method

In this section, the algebraic formulations for efficiently updating the redundancy matrix $\mathbf{R}$ (see Equation (4)) are presented. Three different update scenarios are examined for truss and frame structures: adding, removing, and exchanging elements. Although the exchange of elements is methodically equivalent to removing with subsequently adding elements, the computational cost of a direct exchange is lower. Besides the algebraic formulations, an algorithm is provided for each scenario that focuses on computational efficiency. A MATLAB implementation of all algorithms is publicly available on DaRUS (Krake and von Scheven 2022).

The structure of the derivation is similar for all scenarios and the notation is organized as follows: Symbols without tilde are quantities prior to the update step, whereas symbols with tilde characterize the updated ones (or objects that are new due to the update). Furthermore, due to the generic matrix notation, the derivation covers both truss and frame structures.

### 3.1 Adding elements

Adding a new element to an existing truss or frame (or mixed) structure corresponds to the integration of a new compatibility submatrix $\tilde{\mathbf{a}} \in \mathbb{R}^{\tilde{n}_m \times n}$ into the compatibility matrix $\mathbf{A} \in \mathbb{R}^{n_q \times n}$ and a new material submatrix $\tilde{\mathbf{c}} \in \mathbb{R}^{\tilde{n}_m \times \tilde{n}_m}$ into the material matrix $\mathbf{C} \in \mathbb{R}^{n_q \times n_q}$. To realize this step, we assume that the compatibility matrix $\mathbf{A}$ and material matrix $\mathbf{C}$ are represented by

$$\mathbf{A} = \begin{bmatrix} \mathbf{A}_1 \\ \mathbf{A}_2 \end{bmatrix}, \qquad \mathbf{C} = \begin{bmatrix} \mathbf{C}_1 & \mathbf{0} \\ \mathbf{0} & \mathbf{C}_2 \end{bmatrix}, \tag{6}$$

where $\mathbf{A}_1, \mathbf{A}_2, \mathbf{C}_1$, and $\mathbf{C}_2$ are appropriate matrices. The updated compatibility matrix $\tilde{\mathbf{A}} \in \mathbb{R}^{(n_q + \tilde{n}_m) \times n}$ and updated material matrix $\tilde{\mathbf{C}} \in \mathbb{R}^{(n_q + \tilde{n}_m) \times (n_q + \tilde{n}_m)}$ are therefore given by

$$\tilde{\mathbf{A}} = \begin{bmatrix} \mathbf{A}_1 \\ \tilde{\mathbf{a}} \\ \mathbf{A}_2 \end{bmatrix}, \qquad \tilde{\mathbf{C}} = \begin{bmatrix} \mathbf{C}_1 & 0 & 0 \\ 0 & \tilde{\mathbf{c}} & 0 \\ 0 & 0 & \mathbf{C}_2 \end{bmatrix}, \tag{7}$$

provided that the integration of the submatrices is performed at the same row index (for a more compact representation, we recommend adding the submatrices below the last row or column, respectively). Consequently, the updated dimensions are given by $\tilde{n}_q = n_q + \tilde{n}_m$ and $\tilde{n} = n$. The next step is to update the elastic stiffness matrix $\mathbf{K} \in \mathbb{R}^{n \times n}$ (see Equation (2)), which is given by

$$\mathbf{K} = \mathbf{A}^\top \mathbf{C} \mathbf{A} = \begin{bmatrix} \mathbf{A}_1 \\ \mathbf{A}_2 \end{bmatrix}^\top \begin{bmatrix} \mathbf{C}_1 & 0 \\ 0 & \mathbf{C}_2 \end{bmatrix} \begin{bmatrix} \mathbf{A}_1 \\ \mathbf{A}_2 \end{bmatrix} = \mathbf{A}_1^\top \mathbf{C}_1 \mathbf{A}_1 + \mathbf{A}_2^\top \mathbf{C}_2 \mathbf{A}_2. \tag{8}$$

The updated elastic stiffness matrix $\tilde{\mathbf{K}} \in \mathbb{R}^{n \times n}$ can now be expressed as a sum of the elastic stiffness matrix $\mathbf{K}$ and the new submatrices $\tilde{\mathbf{a}}$ and $\tilde{\mathbf{c}}$:

$$\tilde{\mathbf{K}} = \tilde{\mathbf{A}}^\top \tilde{\mathbf{C}} \tilde{\mathbf{A}} = \begin{bmatrix} \mathbf{A}_1 \\ \tilde{\mathbf{a}} \\ \mathbf{A}_2 \end{bmatrix}^\top \begin{bmatrix} \mathbf{C}_1 & 0 & 0 \\ 0 & \tilde{\mathbf{c}} & 0 \\ 0 & 0 & \mathbf{C}_2 \end{bmatrix} \begin{bmatrix} \mathbf{A}_1 \\ \tilde{\mathbf{a}} \\ \mathbf{A}_2 \end{bmatrix} = \mathbf{A}_1^\top \mathbf{C}_1 \mathbf{A}_1 + \mathbf{A}_2^\top \mathbf{C}_2 \mathbf{A}_2 + \tilde{\mathbf{a}}^\top \tilde{\mathbf{c}} \tilde{\mathbf{a}} = \mathbf{K} + \tilde{\mathbf{a}}^\top \tilde{\mathbf{c}} \tilde{\mathbf{a}}. \tag{9}$$

This representation enables the application of the Woodbury formula (see Equation (5)) to obtain the inverse of the updated elastic stiffness matrix:

$$\begin{aligned} \tilde{\mathbf{K}}^{-1} &= (\mathbf{K} + \tilde{\mathbf{a}}^\top \tilde{\mathbf{c}} \tilde{\mathbf{a}})^{-1} = \mathbf{K}^{-1} - \mathbf{K}^{-1} \tilde{\mathbf{a}}^\top \tilde{\mathbf{c}} (\mathbf{I} + \tilde{\mathbf{a}} \mathbf{K}^{-1} \tilde{\mathbf{a}}^\top \tilde{\mathbf{c}})^{-1} \tilde{\mathbf{a}} \mathbf{K}^{-1} \\ &= \mathbf{K}^{-1} - \mathbf{K}^{-1} \tilde{\mathbf{a}}^\top \tilde{\mathbf{c}} \mathbf{G}^{-1} \tilde{\mathbf{a}} \mathbf{K}^{-1}, \end{aligned} \tag{10}$$

where we define $\mathbf{G}^{-1} = (\mathbf{I} + \tilde{\mathbf{a}} \mathbf{K}^{-1} \tilde{\mathbf{a}}^\top \tilde{\mathbf{c}})^{-1} \in \mathbb{R}^{\tilde{n}_m \times \tilde{n}_m}$. From a mechanical point of view, the inverse $\tilde{\mathbf{K}}^{-1}$ always exists, because adding elements to a kinematically determinate structure always leads to a new kinematically determinate structure. Mathematically, the inverse $\tilde{\mathbf{K}}^{-1}$ exists if and only if $\mathbf{G}^{-1}$ exists. And the latter is true because $\mathbf{G}^{-1} = (\mathbf{I} + \tilde{\mathbf{a}} \mathbf{K}^{-1} \tilde{\mathbf{a}}^\top \tilde{\mathbf{c}})^{-1} = ((\tilde{\mathbf{c}}^{-1} + \tilde{\mathbf{a}} \mathbf{K}^{-1} \tilde{\mathbf{a}}^\top) \tilde{\mathbf{c}})^{-1} = \tilde{\mathbf{c}}^{-1} (\tilde{\mathbf{c}}^{-1} + \tilde{\mathbf{a}} \mathbf{K}^{-1} \tilde{\mathbf{a}}^\top)^{-1}$ is invertible due to the fact that $\tilde{\mathbf{c}}$ is invertible (as a diagonal matrix with positive diagonal) and $\tilde{\mathbf{c}}^{-1} + \tilde{\mathbf{a}} \mathbf{K}^{-1} \tilde{\mathbf{a}}^\top$ is invertible (as it is symmetric positive definite).

Now, the final step is to update the redundancy matrix $\mathbf{R} \in \mathbb{R}^{n_q \times n_q}$ (see Equation (4)). To do this, we represent the redundancy matrix in our notation:

$$\begin{aligned} \mathbf{R} = \mathbf{I} - \mathbf{A} \mathbf{K}^{-1} \mathbf{A}^\top \mathbf{C} &= \begin{bmatrix} \mathbf{I} & 0 \\ 0 & \mathbf{I} \end{bmatrix} - \begin{bmatrix} \mathbf{A}_1 \\ \mathbf{A}_2 \end{bmatrix} \mathbf{K}^{-1} \begin{bmatrix} \mathbf{A}_1 \\ \mathbf{A}_2 \end{bmatrix}^\top \begin{bmatrix} \mathbf{C}_1 & 0 \\ 0 & \mathbf{C}_2 \end{bmatrix} \\ &= \begin{bmatrix} \mathbf{I} - \mathbf{A}_1 \mathbf{K}^{-1} \mathbf{A}_1^\top \mathbf{C}_1 & -\mathbf{A}_1 \mathbf{K}^{-1} \mathbf{A}_2^\top \mathbf{C}_2 \\ -\mathbf{A}_2 \mathbf{K}^{-1} \mathbf{A}_1^\top \mathbf{C}_1 & \mathbf{I} - \mathbf{A}_2 \mathbf{K}^{-1} \mathbf{A}_2^\top \mathbf{C}_2 \end{bmatrix} = \begin{bmatrix} \mathbf{R}_1 & \mathbf{R}_2 \\ \mathbf{R}_3 & \mathbf{R}_4 \end{bmatrix}, \end{aligned} \tag{11}$$

where $\mathbf{R}_1, \mathbf{R}_2, \mathbf{R}_3$, and $\mathbf{R}_4$ are appropriate matrices. We use this equation to derive the final formulation of the updated redundancy matrix $\tilde{\mathbf{R}} \in \mathbb{R}^{(n_q + \tilde{n}_m) \times (n_q + \tilde{n}_m)}$:

$$\tilde{\mathbf{R}} = \mathbf{I} - \tilde{\mathbf{A}} \tilde{\mathbf{K}}^{-1} \tilde{\mathbf{A}}^\top \tilde{\mathbf{C}} \tag{12}$$

$$= \mathbf{I} - \tilde{\mathbf{A}} (\mathbf{K}^{-1} - \mathbf{K}^{-1} \tilde{\mathbf{a}}^\top \tilde{\mathbf{c}} \mathbf{G}^{-1} \tilde{\mathbf{a}} \mathbf{K}^{-1}) \tilde{\mathbf{A}}^\top \tilde{\mathbf{C}}$$

$$= \begin{bmatrix} \mathbf{I} & 0 & 0 \\ 0 & \mathbf{I} & 0 \\ 0 & 0 & \mathbf{I} \end{bmatrix} - \begin{bmatrix} \mathbf{A}_1 \mathbf{K}^{-1} \mathbf{A}_1^\top \mathbf{C}_1 & \mathbf{A}_1 \mathbf{K}^{-1} \tilde{\mathbf{a}}^\top \tilde{\mathbf{c}} & \mathbf{A}_1 \mathbf{K}^{-1} \mathbf{A}_2^\top \mathbf{C}_2 \\ \tilde{\mathbf{a}} \mathbf{K}^{-1} \mathbf{A}_1^\top \mathbf{C}_1 & \tilde{\mathbf{a}} \mathbf{K}^{-1} \tilde{\mathbf{a}}^\top \tilde{\mathbf{c}} & \tilde{\mathbf{a}} \mathbf{K}^{-1} \mathbf{A}_2^\top \mathbf{C}_2 \\ \mathbf{A}_2 \mathbf{K}^{-1} \mathbf{A}_1^\top \mathbf{C}_1 & \mathbf{A}_2 \mathbf{K}^{-1} \tilde{\mathbf{a}}^\top \tilde{\mathbf{c}} & \mathbf{A}_2 \mathbf{K}^{-1} \mathbf{A}_2^\top \mathbf{C}_2 \end{bmatrix}$$

$$+ \begin{bmatrix} \mathbf{A}_1 \mathbf{K}^{-1} \tilde{\mathbf{a}}^\top \tilde{\mathbf{c}} \mathbf{G}^{-1} \tilde{\mathbf{a}} \mathbf{K}^{-1} \mathbf{A}_1^\top \mathbf{C}_1 & \mathbf{A}_1 \mathbf{K}^{-1} \tilde{\mathbf{a}}^\top \tilde{\mathbf{c}} \mathbf{G}^{-1} \tilde{\mathbf{a}} \mathbf{K}^{-1} \tilde{\mathbf{a}}^\top \tilde{\mathbf{c}} & \mathbf{A}_1 \mathbf{K}^{-1} \tilde{\mathbf{a}}^\top \tilde{\mathbf{c}} \mathbf{G}^{-1} \tilde{\mathbf{a}} \mathbf{K}^{-1} \mathbf{A}_2^\top \mathbf{C}_2 \\ \tilde{\mathbf{a}} \mathbf{K}^{-1} \tilde{\mathbf{a}}^\top \tilde{\mathbf{c}} \mathbf{G}^{-1} \tilde{\mathbf{a}} \mathbf{K}^{-1} \mathbf{A}_1^\top \mathbf{C}_1 & \tilde{\mathbf{a}} \mathbf{K}^{-1} \tilde{\mathbf{a}}^\top \tilde{\mathbf{c}} \mathbf{G}^{-1} \tilde{\mathbf{a}} \mathbf{K}^{-1} \tilde{\mathbf{a}}^\top \tilde{\mathbf{c}} & \tilde{\mathbf{a}} \mathbf{K}^{-1} \tilde{\mathbf{a}}^\top \tilde{\mathbf{c}} \mathbf{G}^{-1} \tilde{\mathbf{a}} \mathbf{K}^{-1} \mathbf{A}_2^\top \mathbf{C}_2 \\ \mathbf{A}_2 \mathbf{K}^{-1} \tilde{\mathbf{a}}^\top \tilde{\mathbf{c}} \mathbf{G}^{-1} \tilde{\mathbf{a}} \mathbf{K}^{-1} \mathbf{A}_1^\top \mathbf{C}_1 & \mathbf{A}_2 \mathbf{K}^{-1} \tilde{\mathbf{a}}^\top \tilde{\mathbf{c}} \mathbf{G}^{-1} \tilde{\mathbf{a}} \mathbf{K}^{-1} \tilde{\mathbf{a}}^\top \tilde{\mathbf{c}} & \mathbf{A}_2 \mathbf{K}^{-1} \tilde{\mathbf{a}}^\top \tilde{\mathbf{c}} \mathbf{G}^{-1} \tilde{\mathbf{a}} \mathbf{K}^{-1} \mathbf{A}_2^\top \mathbf{C}_2 \end{bmatrix}$$

$$= \begin{bmatrix} \mathbf{R}_1 & 0 & \mathbf{R}_2 \\ 0 & 0 & 0 \\ \mathbf{R}_3 & 0 & \mathbf{R}_4 \end{bmatrix} + \begin{bmatrix} 0 & -\mathbf{A}_1 \mathbf{K}^{-1} \tilde{\mathbf{a}}^\top \tilde{\mathbf{c}} & 0 \\ -\tilde{\mathbf{a}} \mathbf{K}^{-1} \mathbf{A}_1^\top \mathbf{C}_1 & \mathbf{I} - \tilde{\mathbf{a}} \mathbf{K}^{-1} \tilde{\mathbf{a}}^\top \tilde{\mathbf{c}} & -\tilde{\mathbf{a}} \mathbf{K}^{-1} \mathbf{A}_2^\top \mathbf{C}_2 \\ 0 & -\mathbf{A}_2 \mathbf{K}^{-1} \tilde{\mathbf{a}}^\top \tilde{\mathbf{c}} & 0 \end{bmatrix}$$

$$+ \begin{bmatrix} \mathbf{A}_1 \mathbf{K}^{-1} \tilde{\mathbf{a}}^\top \tilde{\mathbf{c}} \mathbf{G}^{-1} \tilde{\mathbf{a}} \mathbf{K}^{-1} \mathbf{A}_1^\top \mathbf{C}_1 & \mathbf{A}_1 \mathbf{K}^{-1} \tilde{\mathbf{a}}^\top \tilde{\mathbf{c}} \mathbf{G}^{-1} \tilde{\mathbf{a}} \mathbf{K}^{-1} \tilde{\mathbf{a}}^\top \tilde{\mathbf{c}} & \mathbf{A}_1 \mathbf{K}^{-1} \tilde{\mathbf{a}}^\top \tilde{\mathbf{c}} \mathbf{G}^{-1} \tilde{\mathbf{a}} \mathbf{K}^{-1} \mathbf{A}_2^\top \mathbf{C}_2 \\ \tilde{\mathbf{a}} \mathbf{K}^{-1} \tilde{\mathbf{a}}^\top \tilde{\mathbf{c}} \mathbf{G}^{-1} \tilde{\mathbf{a}} \mathbf{K}^{-1} \mathbf{A}_1^\top \mathbf{C}_1 & \tilde{\mathbf{a}} \mathbf{K}^{-1} \tilde{\mathbf{a}}^\top \tilde{\mathbf{c}} \mathbf{G}^{-1} \tilde{\mathbf{a}} \mathbf{K}^{-1} \tilde{\mathbf{a}}^\top \tilde{\mathbf{c}} & \tilde{\mathbf{a}} \mathbf{K}^{-1} \tilde{\mathbf{a}}^\top \tilde{\mathbf{c}} \mathbf{G}^{-1} \tilde{\mathbf{a}} \mathbf{K}^{-1} \mathbf{A}_2^\top \mathbf{C}_2 \\ \mathbf{A}_2 \mathbf{K}^{-1} \tilde{\mathbf{a}}^\top \tilde{\mathbf{c}} \mathbf{G}^{-1} \tilde{\mathbf{a}} \mathbf{K}^{-1} \mathbf{A}_1^\top \mathbf{C}_1 & \mathbf{A}_2 \mathbf{K}^{-1} \tilde{\mathbf{a}}^\top \tilde{\mathbf{c}} \mathbf{G}^{-1} \tilde{\mathbf{a}} \mathbf{K}^{-1} \tilde{\mathbf{a}}^\top \tilde{\mathbf{c}} & \mathbf{A}_2 \mathbf{K}^{-1} \tilde{\mathbf{a}}^\top \tilde{\mathbf{c}} \mathbf{G}^{-1} \tilde{\mathbf{a}} \mathbf{K}^{-1} \mathbf{A}_2^\top \mathbf{C}_2 \end{bmatrix}.$$

This formulation emphasizes the connection between the previous old matrix and the updated one. We observe that the updated redundancy matrix consists of three different terms: the old redundancy matrix $\mathbf{R}$ characterized by the parts $\mathbf{R}_1, \ldots, \mathbf{R}_4$ (the splitting simply results from the notation), a sparse matrix that characterizes the new element in the structure, and a dense matrix representing the general change due to the integration of a new element.

```
 1  function EfficientUpdateAdd(A, C, K⁻¹, R, ã, c̃)
 2      Define A₁, A₂, C₁, and C₂                                          ▷ see Equation (6)
 3      Update Ã and C̃                                                    ▷ see Equation (7)
 4      Compute K⁻¹ãᵀ
 5      Compute a(K⁻¹ãᵀ)
 6      Compute (aK⁻¹ãᵀ)c̃
 7      Compute G = I + ãK⁻¹ãᵀc̃
 8      Compute G⁻¹ = (I + ãK⁻¹ãᵀc̃)⁻¹
 9      Compute c̃G⁻¹
10      Update K̃⁻¹ = K⁻¹ − (K⁻¹ãᵀ)(c̃G⁻¹)(K⁻¹ãᵀ)ᵀ                         ▷ see Equation (10)
11      Compute A₁(K⁻¹ãᵀ) and A₂(K⁻¹ãᵀ)
12      Compute (A₁K⁻¹ãᵀ)c̃ and (A₂K⁻¹ãᵀ)c̃
13      Compute ãK⁻¹A₁ᵀC₁ = (A₁K⁻¹ãᵀ)ᵀC₁ and ãK⁻¹A₂ᵀC₂ = (A₂K⁻¹ãᵀ)ᵀC₂
14      Update R̃                                                           ▷ see Equation (12)
15      return Ã, C̃, K̃⁻¹, and R̃
16  end function
```

**Algorithm 1** Efficient update of the redundancy matrix (and related matrices) when an element is added to a truss or frame structure. The process is designed such that it can be performed in a repeatable manner.

To clearly represent the entire update step, Algorithm 1 summarizes the individual update formulas, while focusing on computational efficiency. In general, the update (and thus the algorithm) is designed such that it can be performed in a repeatable manner. This aspect can be observed in Algorithm 1, where the input in line 1 is the same as the output in line 15, except for the new quantities $\tilde{\mathbf{a}}$ and $\tilde{\mathbf{c}}$ (characterizing the new element). The lines in-between describe the update steps of the four relevant matrices: $\mathbf{A}$, $\mathbf{C}$, $\mathbf{K}^{-1}$, and $\mathbf{R}$.

The update of the compatibility matrix $\mathbf{A}$ and material matrix $\mathbf{C}$ is performed in lines 2–3, where the new elements $\tilde{\mathbf{a}}$ and $\tilde{\mathbf{c}}$ are simply inserted at a specific row into the two matrices (see Equation (6) and Equation (7)). In lines 4–9, various quantities are computed successively that are used to update the inverse of the elastic stiffness matrix $\mathbf{K}^{-1}$ in line 10. In the case of truss structures, i.e., $\tilde{n}_\mathrm{m} = 1$ (otherwise it is $\tilde{n}_\mathrm{m} = 3$ or $\tilde{n}_\mathrm{m} = 6$), the computations in lines 4–7 are matrix-vector multiplications and the inverse in line 8 is simply the reciprocal. Moreover, the update formula in line 10 (see Equation (10)) can be easily computed via an outer product. The same applies to the update of the redundancy matrix $\mathbf{R}$, where relevant parts are computed by matrix-vector operations in lines 11–13 and the final update happens in line 14. The update formula can also be computed via outer products (see Equation (12)).

In sum, we observe that many computations can be deduced from previous results and all operations are matrix-vector multiplications, matrix additions, outer products, or thin matrix inverses (of dimension $\tilde{n}_\mathrm{m} \times \tilde{n}_\mathrm{m}$). Consequently, the algorithm has a computational complexity of $O(n_\mathrm{q}^2)$ with a small coefficient. If the entire recomputation is used to update the redundancy matrix, the computational complexity is given by $O(n \cdot n_\mathrm{q}^2)$.

### 3.2 Removing elements

Removing an element $i_0$ from an existing truss or frame (or mixed) structure equals the deletion of specific rows or columns in the compatibility matrix $\mathbf{A} \in \mathbb{R}^{n_\mathrm{q} \times n}$ and material matrix $\mathbf{C} \in \mathbb{R}^{n_\mathrm{q} \times n_\mathrm{q}}$. To this end, we assume that the compatibility matrix $\mathbf{A}$ and material matrix $\mathbf{C}$ are represented by

$$\mathbf{A} = \begin{bmatrix} \mathbf{A}_1 \\ \mathbf{a} \\ \mathbf{A}_2 \end{bmatrix}, \qquad \mathbf{C} = \begin{bmatrix} \mathbf{C}_1 & \mathbf{0} & \mathbf{0} \\ \mathbf{0} & \mathbf{c} & \mathbf{0} \\ \mathbf{0} & \mathbf{0} & \mathbf{C}_2 \end{bmatrix}, \tag{13}$$

where $\mathbf{A}_1$, $\mathbf{A}_2$, $\mathbf{C}_1$, and $\mathbf{C}_2$ are appropriate matrices and $\mathbf{a} \in \mathbb{R}^{n_\mathrm{m} \times n}$ and $\mathbf{c} \in \mathbb{R}^{n_\mathrm{m} \times n_\mathrm{m}}$ correspond to the specific element $i_0$ that should be removed. The updated compatibility matrix $\tilde{\mathbf{A}} \in \mathbb{R}^{(n_\mathrm{q} - n_\mathrm{m}) \times n}$ and updated material matrix $\tilde{\mathbf{C}} \in \mathbb{R}^{(n_q - n_m) \times (n_q - n_m)}$ are therefore given by

$$\tilde{\mathbf{A}} = \mathbf{S}^\top \mathbf{A} = \begin{bmatrix} \mathbf{A}_1 \\ \mathbf{A}_2 \end{bmatrix}, \qquad \tilde{\mathbf{C}} = \mathbf{S}^\top \mathbf{C} \mathbf{S} = \begin{bmatrix} \mathbf{C}_1 & 0 \\ 0 & \mathbf{C}_2 \end{bmatrix}, \tag{14}$$

where the deletion operation is realized by a matrix $\mathbf{S} \in \mathbb{R}^{n_q \times (n_q - n_m)}$ (the matrix notation of $\mathbf{S}$ is used for theoretical purposes only and, for instance, the implementation of $\mathbf{S}^\top \mathbf{A}$ should not be done via matrix-matrix multiplication). Applying the matrix $\mathbf{S}$ from the left corresponds to the deletion of rows regarding the element $i_0$ and from the right to columns regarding element $i_0$. Consequently, the updated dimensions are given by $\tilde{n}_q = n_q - n_m$ and $\tilde{n} = n$. The next step is to update the elastic stiffness matrix $\mathbf{K} \in \mathbb{R}^{n \times n}$ (see Equation (2)), which is given by

$$\mathbf{K} = \mathbf{A}^\top \mathbf{C} \mathbf{A} = \begin{bmatrix} \mathbf{A}_1 \\ \mathbf{a} \\ \mathbf{A}_2 \end{bmatrix}^\top \begin{bmatrix} \mathbf{C}_1 & 0 & 0 \\ 0 & \mathbf{c} & 0 \\ 0 & 0 & \mathbf{C}_2 \end{bmatrix} \begin{bmatrix} \mathbf{A}_1 \\ \mathbf{a} \\ \mathbf{A}_2 \end{bmatrix} = \mathbf{A}_1^\top \mathbf{C}_1 \mathbf{A}_1 + \mathbf{a}^\top \mathbf{c} \mathbf{a} + \mathbf{A}_2^\top \mathbf{C}_2 \mathbf{A}_2. \tag{15}$$

With this, we can express the updated elastic stiffness matrix $\tilde{\mathbf{K}} \in \mathbb{R}^{n \times n}$ by a sum of the elastic stiffness matrix $\mathbf{K}$ and the deleted submatrices $\mathbf{a}$ and $\mathbf{c}$:

$$\tilde{\mathbf{K}} = \tilde{\mathbf{A}}^\top \tilde{\mathbf{C}} \tilde{\mathbf{A}} = \begin{bmatrix} \mathbf{A}_1 \\ \mathbf{A}_2 \end{bmatrix}^\top \begin{bmatrix} \mathbf{C}_1 & 0 \\ 0 & \mathbf{C}_2 \end{bmatrix} \begin{bmatrix} \mathbf{A}_1 \\ \mathbf{A}_2 \end{bmatrix} = \mathbf{A}_1^\top \mathbf{C}_1 \mathbf{A}_1 + \mathbf{A}_2^\top \mathbf{C}_2 \mathbf{A}_2 = \mathbf{K} - \mathbf{a}^\top \mathbf{c} \mathbf{a}. \tag{16}$$

Now, we can apply the Woodbury formula (see Equation (5)) to obtain the inverse of the updated elastic stiffness matrix:

$$\begin{aligned} \tilde{\mathbf{K}}^{-1} &= (\mathbf{K} - \mathbf{a}^\top \mathbf{c} \mathbf{a})^{-1} = \mathbf{K}^{-1} + \mathbf{K}^{-1} \mathbf{a}^\top \mathbf{c} (\mathbf{I} - \mathbf{a} \mathbf{K}^{-1} \mathbf{a}^\top \mathbf{c})^{-1} \mathbf{a} \mathbf{K}^{-1} \\ &= \mathbf{K}^{-1} + \mathbf{K}^{-1} \mathbf{a}^\top \mathbf{c} \mathbf{G}^{-1} \mathbf{a} \mathbf{K}^{-1}, \end{aligned} \tag{17}$$

where we define $\mathbf{G}^{-1} = (\mathbf{I} - \mathbf{a} \mathbf{K}^{-1} \mathbf{a}^\top \mathbf{c})^{-1} \in \mathbb{R}^{n_m \times n_m}$ (we will later see that $\mathbf{G}$ is already a part of the redundancy matrix $\mathbf{R}$).

Finally, we can update the redundancy matrix $\mathbf{R} \in \mathbb{R}^{n_q \times n_q}$ (see Equation (4)). We represent the deleted submatrices by $\mathbf{a} = \mathbf{E}^\top \mathbf{A}$ and $\mathbf{c} = \mathbf{E}^\top \mathbf{C} \mathbf{E}$, where $\mathbf{E} \in \mathbb{R}^{n_q \times n_m}$ is simply the projection onto the entries that correspond to the element $i_0$ (in the case of truss structures, i.e. $n_m = 1$, $\mathbf{E}$ is a canonical unit vector, otherwise, it is a collection of canonical unit vectors). We also use the identities $\mathbf{A}^\top \mathbf{E} \mathbf{E}^\top \mathbf{C} \mathbf{E} = \mathbf{A}^\top \mathbf{C} \mathbf{E}$ and $\mathbf{A}^\top \mathbf{S} \mathbf{S}^\top \mathbf{C} \mathbf{S} = \mathbf{A}^\top \mathbf{C} \mathbf{S}$, which are easy to prove, to derive the formulation of the updated redundancy matrix $\tilde{\mathbf{R}} \in \mathbb{R}^{(n_q - n_m) \times (n_q - n_m)}$:

$$\begin{aligned} \tilde{\mathbf{R}} &= \mathbf{I} - \tilde{\mathbf{A}} \tilde{\mathbf{K}}^{-1} \tilde{\mathbf{A}}^\top \tilde{\mathbf{C}} \\ &= \mathbf{I} - \tilde{\mathbf{A}} (\mathbf{K}^{-1} + \mathbf{K}^{-1} \mathbf{a}^\top \mathbf{c} \mathbf{G}^{-1} \mathbf{a} \mathbf{K}^{-1}) \tilde{\mathbf{A}}^\top \tilde{\mathbf{C}} \\ &= \mathbf{S}^\top \mathbf{I} \mathbf{S} - \mathbf{S}^\top \mathbf{A} \mathbf{K}^{-1} \mathbf{A}^\top \mathbf{S} \mathbf{S}^\top \mathbf{C} \mathbf{S} - \mathbf{S}^\top \mathbf{A} \mathbf{K}^{-1} \mathbf{a}^\top \mathbf{c} \mathbf{G}^{-1} \mathbf{a} \mathbf{K}^{-1} \mathbf{A}^\top \mathbf{S} \mathbf{S}^\top \mathbf{C} \mathbf{S} \\ &= \mathbf{S}^\top \mathbf{I} \mathbf{S} - \mathbf{S}^\top \mathbf{A} \mathbf{K}^{-1} \mathbf{A}^\top \mathbf{C} \mathbf{S} - \mathbf{S}^\top \mathbf{A} \mathbf{K}^{-1} \mathbf{a}^\top \mathbf{c} \mathbf{G}^{-1} \mathbf{a} \mathbf{K}^{-1} \mathbf{A}^\top \mathbf{C} \mathbf{S} \\ &= \mathbf{S}^\top (\mathbf{I} - \mathbf{A} \mathbf{K}^{-1} \mathbf{A}^\top \mathbf{C}) \mathbf{S} - \mathbf{S}^\top \mathbf{A} \mathbf{K}^{-1} \mathbf{a}^\top \mathbf{c} (\mathbf{I} - \mathbf{a} \mathbf{K}^{-1} \mathbf{a}^\top \mathbf{c})^{-1} \mathbf{a} \mathbf{K}^{-1} \mathbf{A}^\top \mathbf{C} \mathbf{S} \\ &= \mathbf{S}^\top \mathbf{R} \mathbf{S} - \mathbf{S}^\top \mathbf{A} \mathbf{K}^{-1} \mathbf{a}^\top \mathbf{c} (\mathbf{I} - \mathbf{a} \mathbf{K}^{-1} \mathbf{a}^\top \mathbf{c})^{-1} \mathbf{a} \mathbf{K}^{-1} \mathbf{A}^\top \mathbf{C} \mathbf{S} \\ &= \mathbf{S}^\top \mathbf{R} \mathbf{S} - \mathbf{S}^\top \mathbf{A} \mathbf{K}^{-1} \mathbf{A}^\top \mathbf{E} \mathbf{E}^\top \mathbf{C} \mathbf{E} (\mathbf{E}^\top \mathbf{I} \mathbf{E} - \mathbf{E}^\top \mathbf{A} \mathbf{K}^{-1} \mathbf{A}^\top \mathbf{E} \mathbf{E}^\top \mathbf{C} \mathbf{E})^{-1} \mathbf{E}^\top \mathbf{A} \mathbf{K}^{-1} \mathbf{A}^\top \mathbf{C} \mathbf{S} \\ &= \mathbf{S}^\top \mathbf{R} \mathbf{S} - \mathbf{S}^\top \mathbf{A} \mathbf{K}^{-1} \mathbf{A}^\top \mathbf{C} \mathbf{E} (\mathbf{E}^\top \mathbf{I} \mathbf{E} - \mathbf{E}^\top \mathbf{A} \mathbf{K}^{-1} \mathbf{A}^\top \mathbf{C} \mathbf{E})^{-1} \mathbf{E}^\top \mathbf{A} \mathbf{K}^{-1} \mathbf{A}^\top \mathbf{C} \mathbf{S} \\ &= \mathbf{S}^\top \mathbf{R} \mathbf{S} - (\mathbf{S}^\top (\mathbf{I} - \mathbf{A} \mathbf{K}^{-1} \mathbf{A}^\top \mathbf{C}) \mathbf{E}) (\mathbf{E}^\top (\mathbf{I} - \mathbf{A} \mathbf{K}^{-1} \mathbf{A}^\top \mathbf{C}) \mathbf{E})^{-1} (\mathbf{E}^\top (\mathbf{I} - \mathbf{A} \mathbf{K}^{-1} \mathbf{A}^\top \mathbf{C}) \mathbf{S}) \\ &= \mathbf{S}^\top \mathbf{R} \mathbf{S} - (\mathbf{S}^\top \mathbf{R} \mathbf{E}) (\mathbf{E}^\top \mathbf{R} \mathbf{E})^{-1} (\mathbf{E}^\top \mathbf{R} \mathbf{S}). \end{aligned} \tag{18}$$

This formulation shows that the updated redundancy matrix consists of two different terms: the old redundancy matrix $\mathbf{R}$ that only needs to be shrunken and a dense matrix representing the general change due to the removing of the element. We also observe that the two terms can be directly computed by the redundancy matrix $\mathbf{R}$ (while the elastic stiffness matrix $\mathbf{K}$ or the deleted submatrices $\mathbf{a}$ or $\mathbf{c}$ are not necessary). In fact, the inverse $(\mathbf{E}^\top \mathbf{R} \mathbf{E})^{-1}$ is equal to $\mathbf{G}^{-1}$ and, therefore, the update will fail if $(\mathbf{E}^\top \mathbf{R} \mathbf{E})^{-1} = \mathbf{G}^{-1}$ (or equivalently $\mathbf{K}^{-1}$) does not exist. This is the case if the element $i_0$ is a statically determinate part of the structure. As a consequence, the new structure would be kinematically indeterminate.

```
1  function EFFICIENTUPDATEREMOVE(A, C, K⁻¹, R, index i₀)
2      Define A₁, a, A₂, C₁, c, and C₂                              ▷ see Equation (13)
3      Update Ã and C̃                                               ▷ see Equation (14)
4      Compute K⁻¹aᵀ
5      Compute G = I – aK⁻¹aᵀc = EᵀRE
6      Compute G⁻¹ = (I – aK⁻¹aᵀc)⁻¹ = (EᵀRE)⁻¹
7      Update K̃⁻¹ = K⁻¹ + (K⁻¹aᵀ)(cG⁻¹)(K⁻¹aᵀ)ᵀ                    ▷ see Equation (17)
8      Update R̃ = SᵀRS – (SᵀRE)(EᵀRE)⁻¹(EᵀRS)                       ▷ see Equation (18)
9      return Ã, C̃, K̃⁻¹, and R̃
10 end function
```

**Algorithm 2** Efficient update of the redundancy matrix (and related matrices) when an element is removed from a truss or frame structure. The process is designed such that it can be performed in a repeatable manner (although the redundancy matrix can be computed without previous steps).

In general, the formulation for $\tilde{\mathbf{R}}$ is consistent with the formula derived by Chen et al. (2010), which is a special case of our formulation considering truss structures, i.e., $n_m = 1$.

To clearly represent the entire update step, Algorithm 2 summarizes the individual update formulas, while focusing on computational efficiency. Analogously to the previous subsection, the update (and thus the algorithm) is designed such that it can be performed in a repeatable manner. Actually, when updating the redundancy matrix $\mathbf{R}$, the previous computations are not necessary, as Equation (18) is only based on $\mathbf{R}$. However, if different update scenarios are mixed (e.g., removing with a subsequent adding), it is necessary to update the matrices $\mathbf{A}$, $\mathbf{C}$, and $\mathbf{K}^{-1}$ as well. Therefore, Algorithm 2 considers these quantities as well.

The update of the compatibility matrix $\mathbf{A}$ and material matrix $\mathbf{C}$ is performed in lines 2–3, where the submatrices $\mathbf{a}$ and $\mathbf{c}$ belonging to the desired element are simply deleted. In lines 4–6, various quantities are computed successively that are used to update the inverse of the elastic stiffness matrix $\mathbf{K}^{-1}$ in line 7. In the case of removing a truss element, i.e., $n_m = 1$ (otherwise it is $n_m = 3$ or $n_m = 6$), the computations in lines 4–5 are matrix-vector multiplications. Moreover, the update formula in line 7 (see Equation (17)) can be easily computed via an outer product. The redundancy matrix $\mathbf{R}$ can be directly updated in line 8 (see Equation (18)). As mentioned before, the computations that use the matrices $\mathbf{S}$ (from the left or right) are simple deletion operations (and should not be implemented via matrix-matrix multiplication).

In sum, we observe that many computations can be deduced from previous results and all operations are matrix-vector multiplications, matrix additions, outer products, thin matrix inverses (of dimension $n_m \times n_m$), or elementary matrix operations (operations with the matrix $\mathbf{S}$ or $\mathbf{E}$). Consequently, the algorithm has a computational complexity of $O(n_q^2)$ with a small coefficient. If the entire recomputation is used to update the redundancy matrix, the computational complexity is given by $O(n \cdot n_q^2)$.

### 3.3 Exchanging elements

Exchanging an element of an existing truss or frame (or mixed) structure is equal to substituting specific rows or columns in the compatibility matrix $\mathbf{A} \in \mathbb{R}^{n_q \times n}$ and material matrix $\mathbf{C} \in \mathbb{R}^{n_q \times n_q}$. For this purpose, we assume that the compatibility matrix $\mathbf{A}$ and material matrix $\mathbf{C}$ read

$$\mathbf{A} = \begin{bmatrix} \mathbf{A}_1 \\ \mathbf{a} \\ \mathbf{A}_2 \end{bmatrix}, \qquad \mathbf{C} = \begin{bmatrix} \mathbf{C}_1 & \mathbf{0} & \mathbf{0} \\ \mathbf{0} & \mathbf{c} & \mathbf{0} \\ \mathbf{0} & \mathbf{0} & \mathbf{C}_2 \end{bmatrix}, \tag{19}$$

where $\mathbf{A}_1$, $\mathbf{A}_2$, $\mathbf{C}_1$, and $\mathbf{C}_2$ are appropriate matrices and $\mathbf{a} \in \mathbb{R}^{n_m \times n}$ and $\mathbf{c} \in \mathbb{R}^{n_m \times n_m}$ correspond to the specific element with index $i_0$ that should be exchanged. To update these matrices, a new compatibility submatrix $\tilde{\mathbf{a}} \in \mathbb{R}^{\tilde{n}_m \times n}$ and new material submatrix $\tilde{\mathbf{c}} \in \mathbb{R}^{\tilde{n}_m \times \tilde{n}_m}$ are considered. Although it is possible to exchange elements with different numbers of load-carrying modes, we restrict our derivation to $\tilde{n}_m = n_m$. Otherwise, the components need to be embedded such that the dimensions match, e.g., via $\mathbf{Sa} = \tilde{\mathbf{a}}$ (compare Section 3.2). The updated compatibility matrix $\tilde{\mathbf{A}} \in \mathbb{R}^{(n_q - n_m + \tilde{n}_m) \times n}$ and updated material matrix $\tilde{\mathbf{C}} \in \mathbb{R}^{(n_q - n_m + \tilde{n}_m) \times (n_q - n_m + \tilde{n}_m)}$ are thus given by

$$\tilde{\mathbf{A}} = \begin{bmatrix} \mathbf{A}_1 \\ \tilde{\mathbf{a}} \\ \mathbf{A}_2 \end{bmatrix}, \qquad \tilde{\mathbf{C}} = \begin{bmatrix} \mathbf{C}_1 & \mathbf{0} & \mathbf{0} \\ \mathbf{0} & \tilde{\mathbf{c}} & \mathbf{0} \\ \mathbf{0} & \mathbf{0} & \mathbf{C}_2 \end{bmatrix}. \tag{20}$$

Consequently, the updated dimensions are given by $\tilde{n}_q = n_q - n_m + \tilde{n}_m = n_q$ and $\tilde{n} = n$. Next, we want to update the elastic stiffness matrix $\mathbf{K} \in \mathbb{R}^{n \times n}$ (see Equation (2)), which is given by

$$\mathbf{K} = \mathbf{A}^\top \mathbf{C} \mathbf{A} = \begin{bmatrix} \mathbf{A}_1 \\ \mathbf{a} \\ \mathbf{A}_2 \end{bmatrix}^\top \begin{bmatrix} \mathbf{C}_1 & \mathbf{0} & \mathbf{0} \\ \mathbf{0} & \mathbf{c} & \mathbf{0} \\ \mathbf{0} & \mathbf{0} & \mathbf{C}_2 \end{bmatrix} \begin{bmatrix} \mathbf{A}_1 \\ \mathbf{a} \\ \mathbf{A}_2 \end{bmatrix} = \mathbf{A}_1^\top \mathbf{C}_1 \mathbf{A}_1 + \mathbf{a}^\top \mathbf{c} \mathbf{a} + \mathbf{A}_2^\top \mathbf{C}_2 \mathbf{A}_2. \tag{21}$$

Now, the updated elastic stiffness matrix $\tilde{\mathbf{K}} \in \mathbb{R}^{n \times n}$ can be represented by a sum of the elastic stiffness matrix $\mathbf{K}$ and all substitution submatrices $\mathbf{a}$, $\mathbf{c}$, $\tilde{\mathbf{a}}$, and $\tilde{\mathbf{c}}$:

$$\tilde{\mathbf{K}} = \tilde{\mathbf{A}}^\top \tilde{\mathbf{C}} \tilde{\mathbf{A}} = \begin{bmatrix} \mathbf{A}_1 \\ \tilde{\mathbf{a}} \\ \mathbf{A}_2 \end{bmatrix}^\top \begin{bmatrix} \mathbf{C}_1 & \mathbf{0} & \mathbf{0} \\ \mathbf{0} & \tilde{\mathbf{c}} & \mathbf{0} \\ \mathbf{0} & \mathbf{0} & \mathbf{C}_2 \end{bmatrix} \begin{bmatrix} \mathbf{A}_1 \\ \tilde{\mathbf{a}} \\ \mathbf{A}_2 \end{bmatrix} = \mathbf{A}_1^\top \mathbf{C}_1 \mathbf{A}_1 + \mathbf{A}_2^\top \mathbf{C}_2 \mathbf{A}_2 + \tilde{\mathbf{a}}^\top \tilde{\mathbf{c}} \tilde{\mathbf{a}} \tag{22}$$

$$= \mathbf{K} + \tilde{\mathbf{a}}^\top \tilde{\mathbf{c}} \tilde{\mathbf{a}} - \mathbf{a}^\top \mathbf{c} \mathbf{a}$$

$$= \mathbf{K} + \begin{bmatrix} \mathbf{a} \\ \tilde{\mathbf{a}} \end{bmatrix}^\top \begin{bmatrix} -\mathbf{c} & \mathbf{0} \\ \mathbf{0} & \tilde{\mathbf{c}} \end{bmatrix} \begin{bmatrix} \mathbf{a} \\ \tilde{\mathbf{a}} \end{bmatrix} = \mathbf{K} + \mathfrak{a}^\top \mathfrak{c} \mathfrak{a},$$

where we define

$$\mathfrak{a} = \begin{bmatrix} \mathbf{a} \\ \tilde{\mathbf{a}} \end{bmatrix} \quad \text{and} \quad \mathfrak{c} = \begin{bmatrix} -\mathbf{c} & \mathbf{0} \\ \mathbf{0} & \tilde{\mathbf{c}} \end{bmatrix}$$

for the sake of transparency. The formulation enables the application of the Woodbury formula (see Equation (5)) to obtain the inverse of the update elastic stiffness matrix:

$$\tilde{\mathbf{K}}^{-1} = (\mathbf{K} + \mathfrak{a}^\top \mathfrak{c} \mathfrak{a})^{-1} = \mathbf{K}^{-1} - \mathbf{K}^{-1} \mathfrak{a}^\top \mathfrak{c} (\mathbf{I} + \mathfrak{a} \mathbf{K}^{-1} \mathfrak{a}^\top \mathfrak{c})^{-1} \mathfrak{a} \mathbf{K}^{-1} \tag{23}$$

$$= \mathbf{K}^{-1} - \mathbf{K}^{-1} \mathfrak{a}^\top \mathfrak{c} \mathbf{G}^{-1} \mathfrak{a} \mathbf{K}^{-1},$$

where we define $\mathbf{G}^{-1} = (\mathbf{I} + \mathfrak{a} \mathbf{K}^{-1} \mathfrak{a}^\top \mathfrak{c})^{-1} \in \mathbb{R}^{n_m + \tilde{n}_m \times n_m + \tilde{n}_m}$.

We conclude the derivation with the update of the redundancy matrix $\mathbf{R} \in \mathbb{R}^{n_q \times n_q}$ (see Equation (4)). To this end, we represent the redundancy matrix in our notation:

$$\mathbf{R} = \mathbf{I} - \mathbf{A} \mathbf{K}^{-1} \mathbf{A}^\top \mathbf{C} = \begin{bmatrix} \mathbf{I} & \mathbf{0} & \mathbf{0} \\ \mathbf{0} & \mathbf{I} & \mathbf{0} \\ \mathbf{0} & \mathbf{0} & \mathbf{I} \end{bmatrix} - \begin{bmatrix} \mathbf{A}_1 \\ \mathbf{a} \\ \mathbf{A}_2 \end{bmatrix} \mathbf{K}^{-1} \begin{bmatrix} \mathbf{A}_1 \\ \mathbf{a} \\ \mathbf{A}_2 \end{bmatrix}^\top \begin{bmatrix} \mathbf{C}_1 & \mathbf{0} & \mathbf{0} \\ \mathbf{0} & \mathbf{c} & \mathbf{0} \\ \mathbf{0} & \mathbf{0} & \mathbf{C}_2 \end{bmatrix} \tag{24}$$

$$= \begin{bmatrix} \mathbf{I} & \mathbf{0} & \mathbf{0} \\ \mathbf{0} & \mathbf{I} & \mathbf{0} \\ \mathbf{0} & \mathbf{0} & \mathbf{I} \end{bmatrix} - \begin{bmatrix} \mathbf{A}_1 \mathbf{K}^{-1} \mathbf{A}_1^\top \mathbf{C}_1 & \mathbf{A}_1 \mathbf{K}^{-1} \mathbf{a}^\top \mathbf{c} & \mathbf{A}_1 \mathbf{K}^{-1} \mathbf{A}_2^\top \mathbf{C}_2 \\ \mathbf{a} \mathbf{K}^{-1} \mathbf{A}_1^\top \mathbf{C}_1 & \mathbf{a} \mathbf{K}^{-1} \mathbf{a}^\top \mathbf{c} & \mathbf{a} \mathbf{K}^{-1} \mathbf{A}_2^\top \mathbf{C}_2 \\ \mathbf{A}_2 \mathbf{K}^{-1} \mathbf{A}_1^\top \mathbf{C}_1 & \mathbf{A}_2 \mathbf{K}^{-1} \mathbf{a}^\top \mathbf{c} & \mathbf{A}_2 \mathbf{K}^{-1} \mathbf{A}_2^\top \mathbf{C}_2 \end{bmatrix}.$$

We use this equation and the definitions $\Delta \mathbf{a} = \tilde{\mathbf{a}} - \mathbf{a}$ and $\Delta \mathbf{c} = \tilde{\mathbf{c}} - \mathbf{c}$ to derive the final formulation of the updated redundancy matrix $\tilde{\mathbf{R}} \in \mathbb{R}^{n_q - n_m + \tilde{n}_m \times n_q - n_m + \tilde{n}_m}$:

$$\tilde{\mathbf{R}} = \mathbf{I} - \tilde{\mathbf{A}} \tilde{\mathbf{K}}^{-1} \tilde{\mathbf{A}}^\top \tilde{\mathbf{C}} \tag{25}$$

$$= \mathbf{I} - \tilde{\mathbf{A}} (\mathbf{K}^{-1} - \mathbf{K}^{-1} \mathfrak{a}^\top \mathfrak{c} \mathbf{G}^{-1} \mathfrak{a} \mathbf{K}^{-1}) \tilde{\mathbf{A}}^\top \tilde{\mathbf{C}}$$

$$= \mathbf{I} - \begin{bmatrix} \mathbf{A}_1 \mathbf{K}^{-1} \mathbf{A}_1^\top \mathbf{C}_1 & \mathbf{A}_1 \mathbf{K}^{-1} \tilde{\mathbf{a}}^\top \tilde{\mathbf{c}} & \mathbf{A}_1 \mathbf{K}^{-1} \mathbf{A}_2^\top \mathbf{C}_2 \\ \tilde{\mathbf{a}} \mathbf{K}^{-1} \mathbf{A}_1^\top \mathbf{C}_1 & \tilde{\mathbf{a}} \mathbf{K}^{-1} \tilde{\mathbf{a}}^\top \tilde{\mathbf{c}} & \tilde{\mathbf{a}} \mathbf{K}^{-1} \mathbf{A}_2^\top \mathbf{C}_2 \\ \mathbf{A}_2 \mathbf{K}^{-1} \mathbf{A}_1^\top \mathbf{C}_1 & \mathbf{A}_2 \mathbf{K}^{-1} \tilde{\mathbf{a}}^\top \tilde{\mathbf{c}} & \mathbf{A}_2 \mathbf{K}^{-1} \mathbf{A}_2^\top \mathbf{C}_2 \end{bmatrix}$$

$$+ \begin{bmatrix} \mathbf{A}_1 \mathbf{K}^{-1} \mathfrak{a}^\top \mathfrak{c} \mathbf{G}^{-1} \mathfrak{a} \mathbf{K}^{-1} \mathbf{A}_1^\top \mathbf{C}_1 & \mathbf{A}_1 \mathbf{K}^{-1} \mathfrak{a}^\top \mathfrak{c} \mathbf{G}^{-1} \mathfrak{a} \mathbf{K}^{-1} \tilde{\mathbf{a}}^\top \tilde{\mathbf{c}} & \mathbf{A}_1 \mathbf{K}^{-1} \mathfrak{a}^\top \mathfrak{c} \mathbf{G}^{-1} \mathfrak{a} \mathbf{K}^{-1} \mathbf{A}_2^\top \mathbf{C}_2 \\ \tilde{\mathbf{a}} \mathbf{K}^{-1} \mathfrak{a}^\top \mathfrak{c} \mathbf{G}^{-1} \mathfrak{a} \mathbf{K}^{-1} \mathbf{A}_1^\top \mathbf{C}_1 & \tilde{\mathbf{a}} \mathbf{K}^{-1} \mathfrak{a}^\top \mathfrak{c} \mathbf{G}^{-1} \mathfrak{a} \mathbf{K}^{-1} \tilde{\mathbf{a}}^\top \tilde{\mathbf{c}} & \tilde{\mathbf{a}} \mathbf{K}^{-1} \mathfrak{a}^\top \mathfrak{c} \mathbf{G}^{-1} \mathfrak{a} \mathbf{K}^{-1} \mathbf{A}_2^\top \mathbf{C}_2 \\ \mathbf{A}_2 \mathbf{K}^{-1} \mathfrak{a}^\top \mathfrak{c} \mathbf{G}^{-1} \mathfrak{a} \mathbf{K}^{-1} \mathbf{A}_1^\top \mathbf{C}_1 & \mathbf{A}_2 \mathbf{K}^{-1} \mathfrak{a}^\top \mathfrak{c} \mathbf{G}^{-1} \mathfrak{a} \mathbf{K}^{-1} \tilde{\mathbf{a}}^\top \tilde{\mathbf{c}} & \mathbf{A}_2 \mathbf{K}^{-1} \mathfrak{a}^\top \mathfrak{c} \mathbf{G}^{-1} \mathfrak{a} \mathbf{K}^{-1} \mathbf{A}_2^\top \mathbf{C}_2 \end{bmatrix}$$

$$= \mathbf{I} - \begin{bmatrix} \mathbf{A}_1 \mathbf{K}^{-1} \mathbf{A}_1^\top \mathbf{C}_1 & \mathbf{A}_1 \mathbf{K}^{-1} (\mathbf{a} + \Delta \mathbf{a})^\top (\mathbf{c} + \Delta \mathbf{c}) & \mathbf{A}_1 \mathbf{K}^{-1} \mathbf{A}_2^\top \mathbf{C}_2 \\ (\mathbf{a} + \Delta \mathbf{a}) \mathbf{K}^{-1} \mathbf{A}_1^\top \mathbf{C}_1 & (\mathbf{a} + \Delta \mathbf{a}) \mathbf{K}^{-1} (\mathbf{a} + \Delta \mathbf{a})^\top (\mathbf{c} + \Delta \mathbf{c}) & (\mathbf{a} + \Delta \mathbf{a}) \mathbf{K}^{-1} \mathbf{A}_2^\top \mathbf{C}_2 \\ \mathbf{A}_2 \mathbf{K}^{-1} \mathbf{A}_1^\top \mathbf{C}_1 & \mathbf{A}_2 \mathbf{K}^{-1} (\mathbf{a} + \Delta \mathbf{a})^\top (\mathbf{c} + \Delta \mathbf{c}) & \mathbf{A}_2 \mathbf{K}^{-1} \mathbf{A}_2^\top \mathbf{C}_2 \end{bmatrix}$$

$$+ \begin{bmatrix} \mathbf{A}_1\mathbf{K}^{-1}\mathfrak{a}^\top\mathfrak{c}\mathbf{G}^{-1}\mathfrak{a}\mathbf{K}^{-1}\mathbf{A}_1^\top\mathbf{C}_1 & \mathbf{A}_1\mathbf{K}^{-1}\mathfrak{a}^\top\mathfrak{c}\mathbf{G}^{-1}\mathfrak{a}\mathbf{K}^{-1}\tilde{\mathbf{a}}^\top\tilde{\mathbf{c}} & \mathbf{A}_1\mathbf{K}^{-1}\mathfrak{a}^\top\mathfrak{c}\mathbf{G}^{-1}\mathfrak{a}\mathbf{K}^{-1}\mathbf{A}_2^\top\mathbf{C}_2 \\ \tilde{\mathbf{a}}\mathbf{K}^{-1}\mathfrak{a}^\top\mathfrak{c}\mathbf{G}^{-1}\mathfrak{a}\mathbf{K}^{-1}\mathbf{A}_1^\top\mathbf{C}_1 & \tilde{\mathbf{a}}\mathbf{K}^{-1}\mathfrak{a}^\top\mathfrak{c}\mathbf{G}^{-1}\mathfrak{a}\mathbf{K}^{-1}\tilde{\mathbf{a}}^\top\tilde{\mathbf{c}} & \tilde{\mathbf{a}}\mathbf{K}^{-1}\mathfrak{a}^\top\mathfrak{c}\mathbf{G}^{-1}\mathfrak{a}\mathbf{K}^{-1}\mathbf{A}_2^\top\mathbf{C}_2 \\ \mathbf{A}_2\mathbf{K}^{-1}\mathfrak{a}^\top\mathfrak{c}\mathbf{G}^{-1}\mathfrak{a}\mathbf{K}^{-1}\mathbf{A}_1^\top\mathbf{C}_1 & \mathbf{A}_2\mathbf{K}^{-1}\mathfrak{a}^\top\mathfrak{c}\mathbf{G}^{-1}\mathfrak{a}\mathbf{K}^{-1}\tilde{\mathbf{a}}^\top\tilde{\mathbf{c}} & \mathbf{A}_2\mathbf{K}^{-1}\mathfrak{a}^\top\mathfrak{c}\mathbf{G}^{-1}\mathfrak{a}\mathbf{K}^{-1}\mathbf{A}_2^\top\mathbf{C}_2 \end{bmatrix}$$

$$= \mathbf{R} - \begin{bmatrix} \mathbf{0} & \mathbf{A}_1\mathbf{K}^{-1}(\mathbf{a}^\top\Delta\mathbf{c} + \Delta\mathbf{a}^\top\mathbf{c} + \Delta\mathbf{a}^\top\Delta\mathbf{c}) & \mathbf{0} \\ \Delta\mathbf{a}\mathbf{K}^{-1}\mathbf{A}_1^\top\mathbf{C}_1 & \Delta\mathbf{a}\mathbf{K}^{-1}\mathbf{a}^\top\mathbf{c} + \tilde{\mathbf{a}}\mathbf{K}^{-1}(\mathbf{a}^\top\Delta\mathbf{c} + \Delta\mathbf{a}^\top\mathbf{c} + \Delta\mathbf{a}^\top\Delta\mathbf{c}) & \Delta\mathbf{a}\mathbf{K}^{-1}\mathbf{A}_2^\top\mathbf{C}_2 \\ \mathbf{0} & \mathbf{A}_2\mathbf{K}^{-1}(\mathbf{a}^\top\Delta\mathbf{c} + \Delta\mathbf{a}^\top\mathbf{c} + \Delta\mathbf{a}^\top\Delta\mathbf{c}) & \mathbf{0} \end{bmatrix}$$

$$+ \begin{bmatrix} \mathbf{A}_1\mathbf{K}^{-1}\mathfrak{a}^\top\mathfrak{c}\mathbf{G}^{-1}\mathfrak{a}\mathbf{K}^{-1}\mathbf{A}_1^\top\mathbf{C}_1 & \mathbf{A}_1\mathbf{K}^{-1}\mathfrak{a}^\top\mathfrak{c}\mathbf{G}^{-1}\mathfrak{a}\mathbf{K}^{-1}\tilde{\mathbf{a}}^\top\tilde{\mathbf{c}} & \mathbf{A}_1\mathbf{K}^{-1}\mathfrak{a}^\top\mathfrak{c}\mathbf{G}^{-1}\mathfrak{a}\mathbf{K}^{-1}\mathbf{A}_2^\top\mathbf{C}_2 \\ \tilde{\mathbf{a}}\mathbf{K}^{-1}\mathfrak{a}^\top\mathfrak{c}\mathbf{G}^{-1}\mathfrak{a}\mathbf{K}^{-1}\mathbf{A}_1^\top\mathbf{C}_1 & \tilde{\mathbf{a}}\mathbf{K}^{-1}\mathfrak{a}^\top\mathfrak{c}\mathbf{G}^{-1}\mathfrak{a}\mathbf{K}^{-1}\tilde{\mathbf{a}}^\top\tilde{\mathbf{c}} & \tilde{\mathbf{a}}\mathbf{K}^{-1}\mathfrak{a}^\top\mathfrak{c}\mathbf{G}^{-1}\mathfrak{a}\mathbf{K}^{-1}\mathbf{A}_2^\top\mathbf{C}_2 \\ \mathbf{A}_2\mathbf{K}^{-1}\mathfrak{a}^\top\mathfrak{c}\mathbf{G}^{-1}\mathfrak{a}\mathbf{K}^{-1}\mathbf{A}_1^\top\mathbf{C}_1 & \mathbf{A}_2\mathbf{K}^{-1}\mathfrak{a}^\top\mathfrak{c}\mathbf{G}^{-1}\mathfrak{a}\mathbf{K}^{-1}\tilde{\mathbf{a}}^\top\tilde{\mathbf{c}} & \mathbf{A}_2\mathbf{K}^{-1}\mathfrak{a}^\top\mathfrak{c}\mathbf{G}^{-1}\mathfrak{a}\mathbf{K}^{-1}\mathbf{A}_2^\top\mathbf{C}_2 \end{bmatrix}.$$

This formulation emphasizes the connection between the old redundancy matrix and the updated one. We observe that the updated redundancy matrix consists of three different terms: the old redundancy matrix $\mathbf{R}$, a sparse matrix that characterizes the exchanged element in the structure, and a dense matrix representing the general change due to the exchange of an element.

```
 1  function EFFICIENTUPDATEEXCHANGE(A, C, K⁻¹, R, index i₀, ã, c̃)
 2      Define A₁, a, A₂, C₁, c, and C₂                                        ▷ see Equation (19)
 3      Update Ã and C̃                                                         ▷ see Equation (20)
 4      Define
                 𝔞 = [a; ã]   and   𝔠 = [-c  0; 0  c̃]
 5      Compute K⁻¹𝔞⊤
 6      Compute 𝔞(K⁻¹𝔞⊤) via a(K⁻¹𝔞⊤) and ã(K⁻¹𝔞⊤)
 7      Compute (𝔞K⁻¹𝔞⊤)𝔠 via (𝔞K⁻¹𝔞⊤)c and (𝔞K⁻¹𝔞⊤)c̃
 8      Compute G = I + 𝔞K⁻¹𝔞⊤𝔠
 9      Compute G⁻¹ = (I + 𝔞K⁻¹𝔞⊤𝔠)⁻¹
10      Compute 𝔠G⁻¹
11      Update K̃⁻¹ = K⁻¹ − (K⁻¹𝔞⊤)(𝔠G⁻¹)(K⁻¹𝔞⊤)⊤.                              ▷ see Equation (23)
12      Compute A₁(K⁻¹𝔞⊤) and A₂(K⁻¹𝔞⊤)
13      Compute 𝔞K⁻¹A₁⊤C₁ = (A₁K⁻¹𝔞⊤)⊤C₁ and 𝔞K⁻¹A₂⊤C₂ = (A₂K⁻¹𝔞⊤)⊤C₂
14      Compute Δ𝔞K⁻¹A₁⊤C₁ = [−I  I](𝔞K⁻¹A₁⊤C₁), Δ𝔞K⁻¹A₂⊤C₂ = [−I  I](𝔞K⁻¹A₂⊤C₂), and
                 Δ𝔞K⁻¹a⊤c = [−I  I](𝔞K⁻¹a⊤c) = [−I  I](𝔞K⁻¹𝔞⊤𝔠)[−I; 0]
15      Compute δ = a⊤Δc + Δa⊤c + Δa⊤Δc and K⁻¹δ
16      Compute A₁(K⁻¹δ), A₂(K⁻¹δ), and ã(K⁻¹δ)
17      Update R̃                                                              ▷ see Equation (25)
18      return Ã, C̃, K̃⁻¹, and R̃
19  end function
```

**Algorithm 3** Efficient update of the redundancy matrix (and related matrices) when an element is exchanged in a truss or frame structure. The process is designed such that it can be performed in a repeatable manner. Although the exchange of elements is methodically equivalent to removing with subsequently adding of elements, the computational cost of a direct exchange is lower.

To clearly represent the entire update step, Algorithm 3 summarizes the individual update formulas, while focusing on computational efficiency. In general, the update (and thus the algorithm) is designed such that it can be performed in a repeatable manner. In Algorithm 3, the four relevant matrices $\mathbf{A}$, $\mathbf{C}$, $\mathbf{K}^{-1}$, and $\mathbf{R}$ get updated while additionally an index $i_0$ (which determines the element that should be exchanged) and new quantities $\tilde{\mathbf{a}}$ and $\tilde{\mathbf{c}}$ are given.

The update of the compatibility matrix $\mathbf{A}$ and material matrix $\mathbf{C}$ is performed in lines 2–3, where the new quantities $\tilde{\mathbf{a}}$ and $\tilde{\mathbf{c}}$ overwrite the quantities $\mathbf{a}$ and $\mathbf{c}$ (see Equation (19) and Equation (20)). In lines 4–10, various quantities are defined and computed successively that are used to update the inverse of the elastic stiffness matrix $\mathbf{K}^{-1}$ in line 11. In the case of truss structures, i.e., $\tilde{n}_m = 1$ (otherwise it is $\tilde{n}_m = 3$ or $\tilde{n}_m = 6$), the computations in lines 5–8 are common matrix-vector multiplications and the inverse in line 9 is simply the reciprocal. Moreover, the update formula in line 11 (see Equation (23)) can be easily computed via an outer product. The same applies to the update of the redundancy matrix $\mathbf{R}$, where relevant parts are computed by matrix-vector operations in lines 12–17 and the final update happens in line 18. The update formula can also be computed via outer products (see Equation (24)).

We again observe that many computations can be deduced from previous results and all operations are matrix-vector multiplications, matrix additions, outer products, or thin matrix inverses (of dimension $\tilde{n}_m \times \tilde{n}_m$). Consequently, the algorithm has a computational complexity of $O(n_q^2)$. If the entire recomputation is used to update the redundancy matrix, the computational complexity is given by $O(n \cdot n_q^2)$.

## 4 Examples

This section demonstrates the application of the proposed update formulations. The first two examples show the range of application with possible scenarios and provide initial interpretations. On the other hand, the third example has a scaleable size and is used for performance tests.

### 4.1 Introductory Example

In this introductory example, three plane truss structures, as shown in Figure 2, will be examined. Young's modulus $E$ and cross sectional area $A$ are constant for all elements. Starting from structure A, system B can be obtained by adding the new element 3. Removing then element 4 yields structure C. By exchanging element 3 from the connection between nodes 3 and 2 to the connection of nodes 4 and 2, system A is reproduced. The changes to the redundancy matrix will be calculated by the update algorithms described in Section 3, starting from the redundancy matrix for system A.

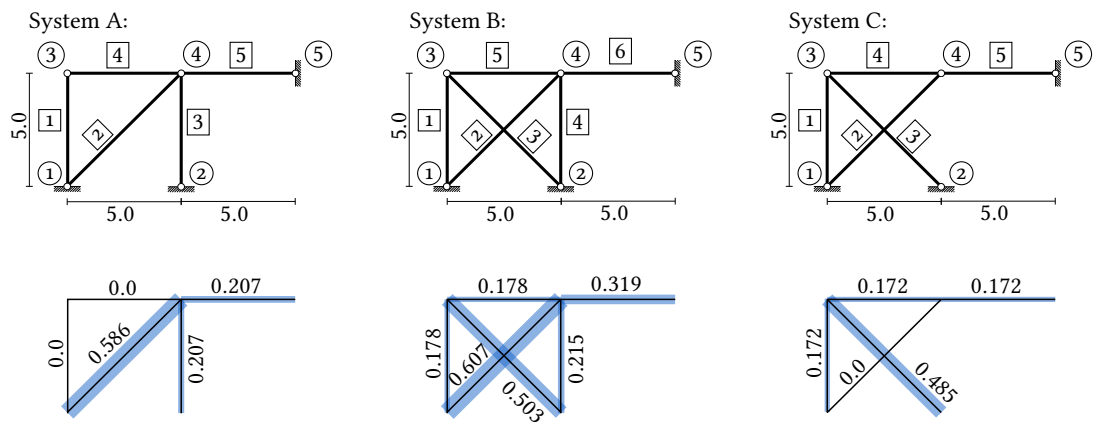

**Figure 2** Introductory example: Three truss systems A, B, and C that result from each other by a single update step (top) and the corresponding redundancy distributions (bottom).

#### 4.1.1 Adding a New Element

Firstly, the modification of the redundancy matrix from system A to system B due to adding a second diagonal element (new element 3) will be demonstrated in detail. Following Algorithm 1, the compatibility matrix $\mathbf{A}$, material matrix $\mathbf{C}$, and redundancy matrix $\mathbf{R}$ of the original system (system A),

$$\mathbf{A} = \begin{bmatrix} 0 & 1 & 0 & 0 \\ 0 & 0 & \sqrt{2}/2 & \sqrt{2}/2 \\ 0 & 0 & 0 & 1 \\ -1 & 0 & 1 & 0 \\ 0 & 0 & -1 & 0 \end{bmatrix}, \mathbf{C} = \operatorname{diag}\begin{bmatrix} 200 \\ 100\sqrt{2} \\ 200 \\ 200 \\ 200 \end{bmatrix}, \mathbf{R} = \begin{bmatrix} 0.0 & 0.0 & 0.0 & 0.0 & 0.0 \\ 0.0 & 0.586 & -0.414 & 0.0 & 0.414 \\ 0.0 & -0.293 & 0.207 & 0.0 & -0.207 \\ 0.0 & 0.0 & 0.0 & 0.0 & 0.0 \\ 0.0 & 0.293 & -0.207 & 0.0 & 0.207 \end{bmatrix} \quad (26)$$

are required as input, as well as the inverse of the stiffness matrix $\mathbf{K} = \mathbf{A}^\top \mathbf{C} \mathbf{A}$.

To add a new element between nodes 2 and 3, the new row $\tilde{\mathbf{a}} = [-\sqrt{2}/2 \; \sqrt{2}/2 \; 0 \; 0]$ of the compatibility matrix and the new entry $\tilde{\mathbf{c}} = [100\sqrt{2}]$ of the material matrix are needed.

Following the steps in Algorithm 1, the redundancy matrix $\tilde{\mathbf{R}}$ for system B can be computed as an update of the redundancy matrix of system A:

$$\tilde{\mathbf{R}} = \begin{bmatrix} 0.178 & -0.0521 & -0.252 & 0.0368 & 0.178 & 0.141 \\ -0.0737 & 0.607 & 0.104 & -0.429 & -0.0737 & 0.356 \\ -0.356 & 0.104 & 0.503 & -0.0737 & -0.356 & -0.282 \\ 0.0368 & -0.304 & -0.0521 & 0.215 & 0.0368 & -0.178 \\ 0.178 & -0.0521 & -0.252 & 0.0368 & 0.178 & 0.141 \\ 0.141 & 0.252 & -0.199 & -0.178 & 0.141 & 0.319 \end{bmatrix}. \quad (27)$$

Besides computational efficiency, the advantage of the presented update algorithms is the possibility to predict the changes to the redundancy matrix a new element will make. One common application could be to avoid bars with zero redundancy, i.e., statically determinate parts in the structure. The modification to the redundancy of existing elements caused by adding a new element is determined only by the diagonal blocks of the last summand in Equation (12), e.g. for the left upper block:

$$\Delta \mathbf{R}_1 = \mathbf{A}_1 \mathbf{K}^{-1} \tilde{\mathbf{a}}^\top \tilde{\mathbf{c}} \mathbf{G}^{-1} \tilde{\mathbf{a}} \mathbf{K}^{-1} \mathbf{A}_1^\top \mathbf{C}_1. \quad (28)$$

If we try to achieve non-zero entries on the diagonals of this matrix $\Delta \mathbf{R}_{11}$, the scalar (for $n_m = 1$) factors $\tilde{\mathbf{c}}$ and $\mathbf{G}^{-1}$ as well as the diagonal matrix $\mathbf{C}_1$ can be neglected for the further derivation. The remaining term is the Gramian matrix of $\tilde{\mathbf{a}} \mathbf{K}^{-1} \mathbf{A}_1^\top$. The term $\mathbf{K}^{-1} \tilde{\mathbf{a}}^\top$ is proportional to the nodal displacements caused by an elongation of the new element, while $\mathbf{A}_1 \mathbf{K}^{-1} \tilde{\mathbf{a}}^\top$ is proportional to the resulting elastic elongations and therefore normal forces in the elements that correspond to $\mathbf{A}_1$. To obtain a non-zero value on a diagonal entry of $\Delta \mathbf{R}_{11}$, the corresponding entry in $\mathbf{A}_1 \mathbf{K}^{-1} \tilde{\mathbf{a}}^\top$ must be non-zero, i.e., an elongation of the new element must cause a normal force in the respective element.

The diagonal terms of $\Delta \mathbf{R}_{11}$ are always non-negative. Mechanically, this means that the redundancy of all elements in the structure can only be increased by adding a new element. For $n_m = 1$, this can be directly seen from Equation (28). The term $\tilde{\mathbf{c}} \mathbf{G}^{-1}$ is a positive scalar and $\mathbf{C}_1$ a diagonal matrix with only positive entries. The remaining term is the Gramian matrix of $\tilde{\mathbf{a}} \mathbf{K}^{-1} \mathbf{A}_1^\top$, which has always non-negative diagonal entries. Cases with $n_m > 1$ can be interpreted as multiple applications of the update algorithm with $n_m = 1$ and therefore follow the same rule.

### 4.1.2 Removing an Element

In the second step, element 4 will be removed from system B to obtain system C. Again, an update procedure as described in Section 3.2 and Algorithm 2 will be used instead of recalculating the redundancy matrix from scratch. Starting point is system B with its compatibility matrix $\mathbf{A}$ and material matrix $\mathbf{C}$,

$$\mathbf{A} = \begin{bmatrix} 0 & 1 & 0 & 0 \\ 0 & 0 & \sqrt{2}/2 & \sqrt{2}/2 \\ -\sqrt{2}/2 & \sqrt{2}/2 & 0 & 0 \\ 0 & 0 & 0 & 1 \\ -1 & 0 & 1 & 0 \\ 0 & 0 & -1 & 0 \end{bmatrix} \qquad \mathbf{C} = \mathrm{diag}\begin{bmatrix} 200 \\ 100\sqrt{2} \\ 100\sqrt{2} \\ 200 \\ 200 \\ 200 \end{bmatrix}, \quad (29)$$

as well as the inverse of the stiffness matrix $\mathbf{K} = \mathbf{A}^\top \mathbf{C} \mathbf{A}$, and the redundancy matrix $\mathbf{R}$ as given in Equation (27) as $\tilde{\mathbf{R}}$.

Following Algorithm 2, first the inverse of the stiffness matrix $\mathbf{K}$ can be updated according to Equation (17). The redundancy matrix itself is updated as described in the last line of Equation (18) or line 10 in Algorithm 2. For this example, the matrix $\mathbf{S}$ represents the deletion of the fourth row or column. This yields the updated redundancy matrix of system C as:

$$\tilde{\mathbf{R}} = \begin{bmatrix} 0.172 & 0.0 & -0.243 & 0.172 & 0.172 \\ 0.0 & 0.0 & 0.0 & 0.0 & 0.0 \\ -0.343 & 0.0 & 0.485 & -0.343 & -0.343 \\ 0.172 & 0.0 & -0.243 & 0.172 & 0.172 \\ 0.172 & 0.0 & -0.243 & 0.172 & 0.172 \end{bmatrix}. \quad (30)$$

The modification to the redundancy of existing elements caused by removing an element is determined by the second term in the last line of Equation (18):

$$\Delta \mathbf{R} = (\mathbf{S}^T \mathbf{R} \mathbf{E})(\mathbf{E}^T \mathbf{R} \mathbf{E})^{-1}(\mathbf{E}^T \mathbf{R} \mathbf{S}). \tag{31}$$

For $n_m = 1$, the term $(\mathbf{E}^T \mathbf{R} \mathbf{E})$ is a scalar, equal to the redundancy of the element to be removed and therefore always positive. The remaining term $(\mathbf{S}^T \mathbf{R} \mathbf{E})(\mathbf{E}^T \mathbf{R} \mathbf{S})$ is the Gramian matrix of $\mathbf{E}^T \mathbf{R} \mathbf{S}$ and therefore has non-negative diagonal entries. Hence, the diagonal terms of $\Delta \mathbf{R}$ are all non-negative. Mechanically this means, that the redundancy of remaining elements in the structure can only be reduced by removing an element. This statement also applies to cases with $n_m > 1$, since these can be interpreted as multiple applications of the update algorithm with $n_m = 1$.

### 4.1.3 Exchanging an Element

Finally, exchanging element number 3 in system C from nodes 2–3 to nodes 2–4 will reproduce system A. Applying the update procedure as described in Section 3.3 and Algorithm 3, the original redundancy matrix of system A (Equation (26)) should be obtained. The input to the algorithms are the compatibility matrix $\mathbf{A}$ and material matrix $\mathbf{C}$ of system C,

$$\mathbf{A} = \begin{bmatrix} 0 & 1 & 0 & 0 \\ 0 & 0 & \sqrt{2}/2 & \sqrt{2}/2 \\ -\sqrt{2}/2 & \sqrt{2}/2 & 0 & 0 \\ -1 & 0 & 1 & 0 \\ 0 & 0 & -1 & 0 \end{bmatrix}, \qquad \mathbf{C} = \text{diag} \begin{bmatrix} 200 \\ 100\sqrt{2} \\ 100\sqrt{2} \\ 200 \\ 200 \end{bmatrix}, \tag{32}$$

as well as the inverse of the stiffness matrix $\mathbf{K} = \mathbf{A}^T \mathbf{C} \mathbf{A}$ and the redundancy matrix (Equation (30)) of system C. Furthermore, the new row $\tilde{\mathbf{a}} = [0\ 0\ 0\ 1]$ of the compatibility matrix and the new entry $\tilde{\mathbf{c}} = [200]$ of the material matrix corresponding to the new element are required, as well as the position of the element to be exchanged $i_0 = 3$.

Using these quantities, the inverse of the stiffness matrix $\mathbf{K}$ can be updated according to Equation (23) or line 11 in Algorithm 3. The redundancy matrix $\mathbf{R}$ itself is updated as described in the last line Equation (25) or line 17 in Algorithm 3.

$$\tilde{\mathbf{R}} = \begin{bmatrix} 0.0 & 0.0 & 0.0 & 0.0 & 0.0 \\ 0.0 & 0.586 & -0.414 & 0.0 & 0.414 \\ 0.0 & -0.293 & 0.207 & 0.0 & -0.207 \\ 0.0 & 0.0 & 0.0 & 0.0 & 0.0 \\ 0.0 & 0.293 & -0.207 & 0.0 & 0.207 \end{bmatrix} \tag{33}$$

The result is identical to the original redundancy matrix $\mathbf{R}$ of system A as given in Equation (26). This indicates that the three efficient update procedures described above are algebraically correct.

## 4.2 Multi-storey Frame Structures

In this subsection, the update algorithms are applied to the analysis of the load-bearing structure of a multi-storey frame structure. For better comprehensibility, we use a simplified model with only idealised hinges or rigid connections. The original structure A is shown in Figure 3 (left) and consists of a bending resistant core modeled with plane Euler-Bernoulli beam elements and attached statically determinate slab-column systems. Young's modulus $E$, cross sectional area $A$, and moment of inertia $I$ are constant for all elements.

The structure has a high degree of statical indeterminacy given by $n_s = 30$. This statical indeterminacy is concentrated in the core as shown in the redundancy distribution in Figure 3 (right). Due to the three load-bearing mechanism in each beam element (elongation, shear, and bending), i.e. $n_m = 3$, the redundancy of every element can be as high as three. The attached slab-column system has zero redundancy. Failure of only one column leads to a progressive collapse of all stories above (Pearson and Delatte 2005). To avoid such catastrophic failure, the robustness of structures should be increased by avoiding large parts in the structure with zero redundancy.

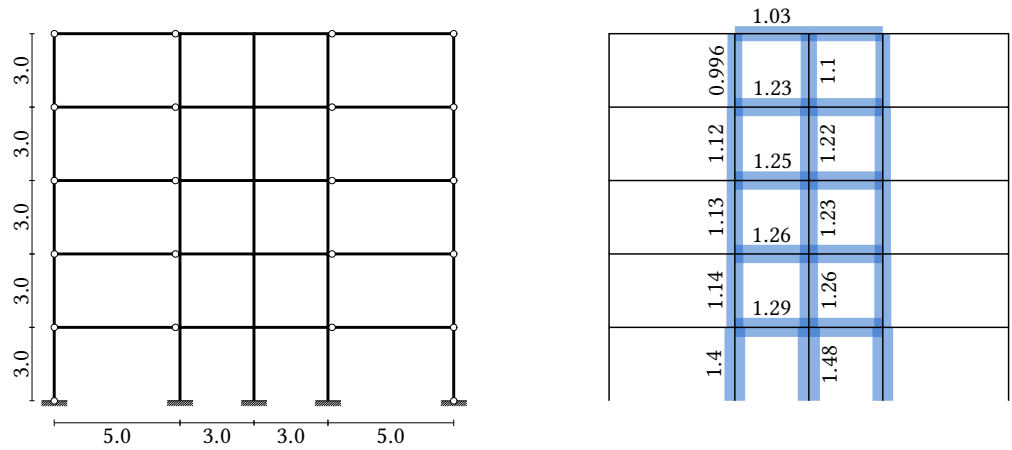

**Figure 3** Multi-storey frame system A with a stiff core and attached statically determinate slab-column systems (left) and the corresponding redundancy distribution (right).

Using the presented efficient update algorithms, the designer can check the redundancy distributions of many variants of a structure to design a robust structure. In this example, two variants of the original structure A will be presented and the necessary updates of the redundancy matrix are described.

Firstly, the redundancy of the slab-column systems is increased by adding diagonal elements as shown in system B (Figure 4, left). The effect of the 10 new truss elements added to the structure on the redundancy matrix can be computed efficiently by applying Algorithm 1 ten times with the respective new compatibility submatrices $\tilde{\mathbf{a}} \in \mathbb{R}^{1 \times n}$. Due to the general matrix notation used for the derivation in Section 3, the updates for all ten elements could also be combined in one step using a combined compatibility submatrix $\tilde{\mathbf{a}} \in \mathbb{R}^{10 \times n}$ and the corresponding material submatrix $\tilde{\mathbf{c}} \in \mathbb{R}^{10 \times 10}$.

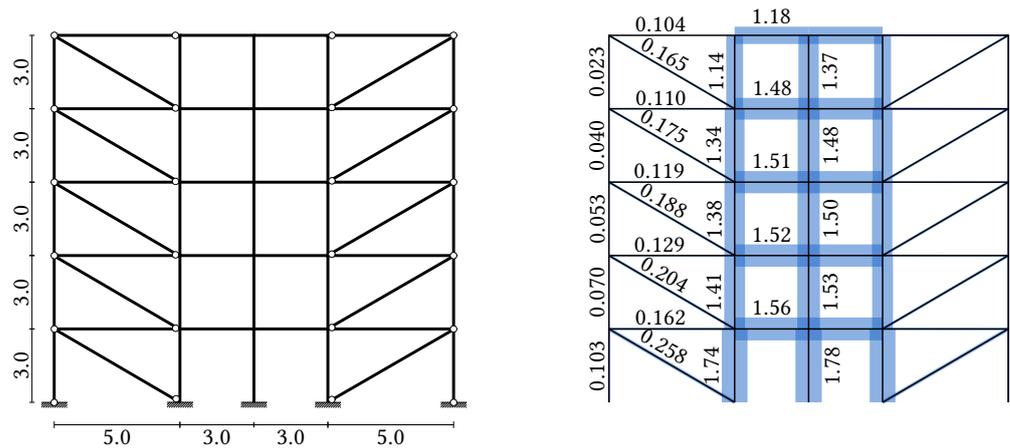

**Figure 4** Multi-storey frame system B with a stiff core and attached statically indeterminate truss systems (left) and the redundancy corresponding distribution (right).

It is obvious that the diagonals make the slab-column systems statically indeterminate. This can be seen in the updated redundancy distribution shown in Figure 4 (right). But still, the redundancy in the indeterminate truss systems is much lower than in the beam elements of the core. This is also due to the higher number of load-bearing mechanisms in the beam elements.

An alternative to adding diagonal elements to the slab-column systems would be, for example, a clamped connection of the slabs and the core. This will also introduce statical indeterminacy to the slab-column part and increase robustness. As shown in system C (Figure 5, left), only the connection between core and the slab is modeled as clamped, whereas the columns are still hinged to the slabs.

To calculate the change of the redundancy matrix due to this modification, various approaches are possible. The intuitive approach would be to remove the truss elements of the slabs using

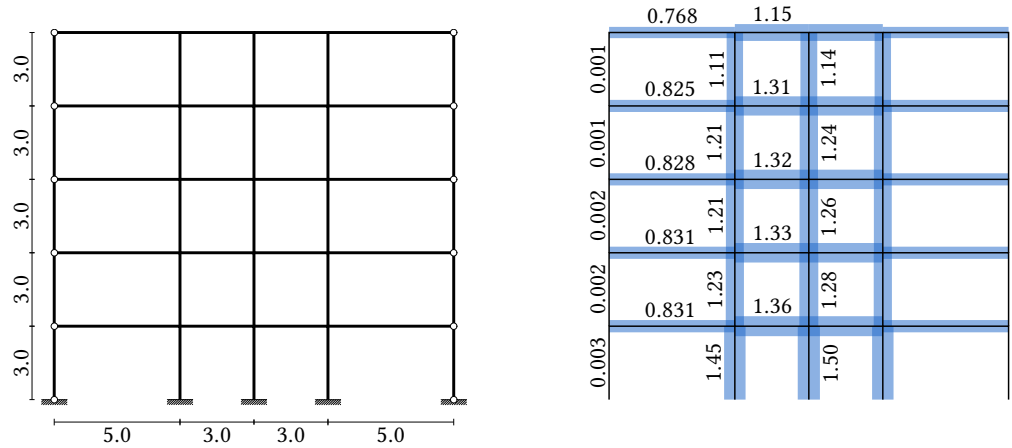

**Figure 5**  Multi-storey frame system C with a stiff core and attached statically indeterminate frame systems (left) and the corresponding redundancy distribution (right).

Algorithm 2 and then add beam elements with Algorithm 1. Due to the general matrix notation used for the derivation in Section 3, the updates for all ten elements can again be combined in one update step. Applying Algorithm 3 for exchanging a truss element with a beam element is not straightforward, as in this case $\tilde{n}_m \neq n_m$. As described in Section 3.3, a deletion or expansion matrix $\mathbf{S}$ needs to be introduced to match the dimensions.

Instead of first removing the truss elements, it is also possible to directly add the two new load-bearing mechanisms of the beam element by adding two rows to the compatibility matrix via $\tilde{\mathbf{a}} \in \mathbb{R}^{2 \times n}$ and Algorithm 1. By avoiding the additional remove and reducing the size of the compatibility submatrix $\tilde{\mathbf{a}}$ for the add, this will increase performance noticeably. Again, the updates for all 10 elements can be combined in one step using a combined compatibility submatrix $\tilde{\mathbf{a}} \in \mathbb{R}^{20 \times n}$.

The redundancy distribution of system C is shown in Figure 5 (right). It can be seen that also for this modification all elements in the system have a non-zero redundancy. However, compared to system B, the redundancy of the columns is very small. This means that, in the case of failure of one column, the displacement of the slabs under self-weight become larger compared to system B (compare with (Bahndorf 1991)). Using the efficient update algorithms described in this paper, the designer can efficiently check the redundancy distributions for different variants of a structure and then decide for the most suitable solution.

### 4.3 Performance Test

In this subsection, we evaluate the efficiency of our update formulas. To this end, we apply our update algorithms to a scaleable truss structure. It is generated by stacking a unit cell $k$ times in all three spacial directions. To get an impression of how the structure scales, a visualization for $k = 1$, $k = 2$, and $k = 3$ is provided in Figure 6. It can be observed that the structure consists of block-like truss structures with additional diagonal elements. For all nodes on the planes $x = 0$, $y = 0$, and $z = 0$, all three degrees of freedom are constrained. Young's modulus $E$ and cross sectional area $A$ are constant for all elements.

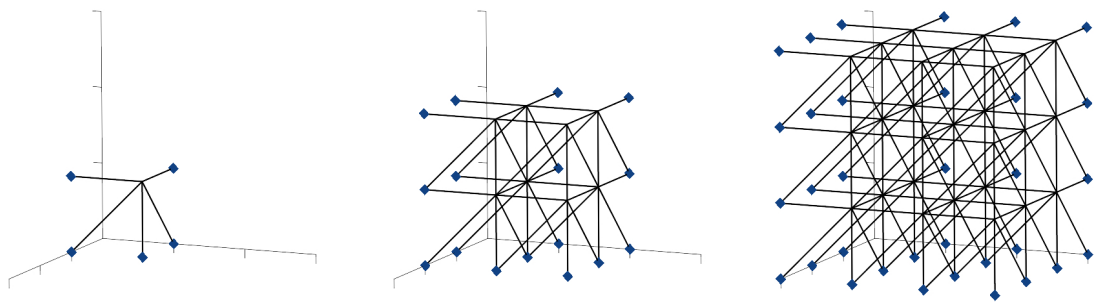

**Figure 6**  Scaleable structure for the parameters $k = 1$, $k = 2$, and $k = 3$.

The statical indeterminacy of the structure depends on the parameter $k$ and is given by $n_s = 2k^3$. This is also the case for the number of elements $n_e = 5k^3$ and the number of degrees of freedom $n = 3k^3$. However, the structure has a fixed $n_e$-$n$-ratio of 5/3. Besides these numbers, Table 1 also shows the averaged computation time for adding, removing, and exchanging a single element with our efficient update formulas. These numbers are compared to the computation time of an entire recomputation (since recomputation is similar for the three scenarios, a single value is used). The evaluation was done with MATLAB R2020b on a machine with 2.90 GHz Intel Core i9-8950HK processor and 32 GB RAM. An implementation of the scaleable truss structure for arbitrary $k$ is also publicly available on DaRUS (Krake and von Scheven 2022).

|  |  |  |  | efficient adding | | efficient removing | | efficient exchanging | |
|---|---|---|---|---|---|---|---|---|---|
| $k$ | $n_e$ | $n$ | recomp. | time | speedup | time | speedup | time | speedup |
| 4 | 320 | 192 | 2 ms | 0 ms | 8.27 | 0 ms | 10.79 | 0 ms | 5.35 |
| 6 | 1080 | 648 | 26 ms | 3 ms | 8.57 | 2 ms | 10.54 | 3 ms | 7.48 |
| 8 | 2560 | 1536 | 217 ms | 11 ms | 19.46 | 10 ms | 21.02 | 17 ms | 12.84 |
| 10 | 5000 | 3000 | 1557 ms | 35 ms | 44.11 | 35 ms | 44.97 | 56 ms | 27.98 |
| 12 | 8640 | 5184 | 8603 ms | 103 ms | 83.36 | 102 ms | 84.06 | 172 ms | 49.99 |
| 14 | 13720 | 8232 | 33201 ms | 262 ms | 126.86 | 272 ms | 122.00 | 451 ms | 73.60 |
| 16 | 20480 | 12288 | 113409 ms | 589 ms | 192.70 | 603 ms | 188.21 | 960 ms | 118.16 |
| 18 | 29160 | 17496 | 306149 ms | 1189 ms | 257.42 | 1143 ms | 267.96 | 1812 ms | 168.94 |
| 20 | 40000 | 24000 | 809161 ms | 2496 ms | 324.24 | 2248 ms | 360.01 | 3537 ms | 228.79 |

**Table 1**   Averaged computation times for the entire recomputation (which is similar for all three scenarios) and our update algorithms. Besides the raw measurements, the speedup ratio is shown.

For all three modifications, it can be observed that our algorithms perform significantly faster. We achieve ratios up to 360 times faster. This relative value grows even more for increasing number of elements. The reason for this is that our algorithm has quadratic computational complexity, whereas the entire recomputation has a cubic one. This fact can be observed in Figure 7, where the computation times are plotted as a function of the number of elements $n_e$. The three efficient update algorithms (adding in green, removing in orange, exchanging in blue) are parallel to the dashed black line, which corresponds to quadratic behavior (the slope is 2). In contrast, the entire recomputation (purple line) is not parallel to the dashed black line. Instead, the recomputation is parallel to the dotted black line, indicating cubic behavior (the slope is 3).

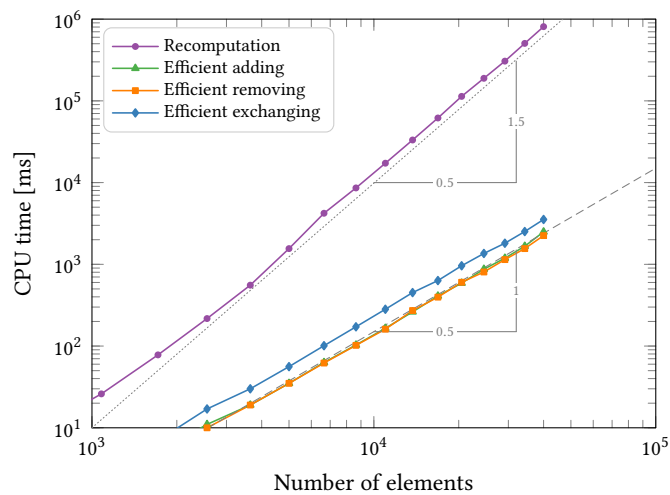

**Figure 7**   Log-log plot for the computation times of the entire recomputation and our update formulas. The computation time is plotted against the number of elements (compare Table 1). The dashed black line corresponds to quadratic behavior, whereas the dotted black line characterizes cubic behavior.

Another important observation is that the update process for structures with a large number of elements still only takes a few seconds, whereas a recomputation takes more than hundreds of seconds. In general, these low absolute values enable an almost instantaneous interaction with a structure, which allows an effective investigation of large buildings via updates.

Besides the comparison to the entire recomputation, we also observe that the exchange operation is faster than a combined removing with subsequent adding. In fact, exchanging takes approximately three-fourths of the time. This aspect should be taken into account if large structures are investigated or/and multiple exchange operations are performed.

## 5 Conclusion

This paper addresses the problem of efficiently updating the redundancy matrix for truss and frame structures. The proposed generic algebraic formulations enable adding, removing, and exchanging of elements, groups of elements, or individual load-carrying types without computing relevant quantities from scratch. As a result, not only computational effort is reduced while preserving analytical correctness and enabling instantaneous interaction with a structure, but also new insights into the update process of statically indeterminate structures are gained. One of these insights is the understanding of how a new element affects certain regions in the structure. In sum, new possibilities for the design and analysis of truss and frame structures are provided.

In the future, the update procedures can be extended to be also applicable for kinematically indeterminate structures or non-linear problems. Furthermore, we plan to exploit the update formulations even more to purposely achieve a specific redundancy distribution in a structure. This may be done either manually by studying cause-and-effect relationships in the update steps, or algorithmically by solving an optimization problem that exploits the efficiency of the update methods. In this regard, the development of an appropriate software tool that guides the user is a relevant research direction.

A completely different idea is to use the formulas to study the inverse problem, i.e., given a redundancy matrix, the goal is to find a suitable structure. In this context, the update formulations may help to find a structure that can be realized mechanically.

**Authors' contributions**    Tim Krake: Conceptualization, Methodology, Software, Visualization, Writing – Original Draft, Writing – Review and Editing. Malte von Scheven: Conceptualization, Software, Visualization, Writing – Original Draft, Writing – Review and Editing. Jan Gade: Conceptualization, Writing – Original Draft, Writing – Review and Editing. Moataz Abdelaal: Conceptualization, Software, Visualization, Writing – Review and Editing. Daniel Weiskopf: Conceptualization, Writing – Review and Editing, Supervision. Manfred Bischoff: Conceptualization, Writing – Review and Editing, Supervision.

**Supplementary Material**    Matlab files: 10.18419/darus-2870

**Acknowledgements**    This work is partly funded by "Deutsche Forschungsgemeinschaft" (DFG, German Research Foundation)—Project-ID 251654672–TRR 161; Project-ID 279064222–SFB 1244; and under Germany's Excellence Strategy EXC 2120/1–390831618.

**Ethics approval and consent to participate**    Not applicable.

**Consent for publication**    Not applicable.

**Competing interests**    The authors declare that they have no competing interests.

**Journal's Note**    JTCAM remains neutral with regard to the content of the publication and institutional affiliations.